\documentclass[fleqn,usenatbib]{mnras}

\usepackage{newtxtext,newtxmath}

\usepackage[T1]{fontenc}

\DeclareRobustCommand{\VAN}[3]{#2}
\let\VANthebibliography\thebibliography
\def\thebibliography{\DeclareRobustCommand{\VAN}[3]{##3}\VANthebibliography}

\usepackage{graphicx}	
\usepackage{amsmath}	

\newcommand{\swift}{\textsc{swift}}
\newcommand{\sphenix}{\textsc{sphenix}}
\newcommand{\eagle}{\textsc{eagle}}
\newcommand{\bahamas}{\textsc{bahamas}}
\newcommand{\msun}{\mathrm{M}_\odot}
\newcommand{\mstar}{M_\star}

\title[BH repositioning]{The importance of black hole repositioning for galaxy formation simulations}

\author[Y. M. Bah\'{e} et al.]{Yannick M. Bah\'{e},$^{1}$\thanks{E-mail: bahe@strw.leidenuniv.nl (YMB)}
Joop Schaye,$^{1}$
Matthieu Schaller,$^{2, 1}$
Richard G.~Bower,$^{3}$
Josh Borrow,$^{3, 4}$
\newauthor
Evgenii Chaikin,$^{1}$
Roi Kugel,$^{1}$
Folkert Nobels$^{1}$
and Sylvia Ploeckinger$^{2}$
\\
$^{1}$Leiden Observatory, Leiden University, PO Box 9513, 2300 RA Leiden, The Netherlands\\
$^{2}$Lorentz Institute for Theoretical Physics, Leiden University, PO Box 9506, 2300 RA Leiden, The Netherlands\\
$^{3}$Institute for Computational Cosmology, Department of Physics, University of Durham, South Road, Durham, DH1 3LE, UK\\
$^{4}$Department of Physics, Kavli Institute for Astrophysics and Space Research, Massachusetts Institute of Technology, Cambridge, MA 02139, USA\\
}

\date{Accepted 2022 May 2. Received 2022 April 21; in original form 2021 September 3}

\pubyear{2022}

\begin{document}
\label{firstpage}
\pagerange{\pageref{firstpage}--\pageref{lastpage}}
\maketitle

\begin{abstract}
Active galactic nucleus (AGN) feedback from accreting supermassive black holes (SMBHs) is an essential ingredient of galaxy formation simulations. The orbital evolution of SMBHs is affected by dynamical friction that cannot be predicted self-consistently by contemporary simulations of galaxy formation in representative volumes. Instead, such simulations typically use a simple ``repositioning'' of SMBHs, but the effects of this approach on SMBH and galaxy properties have not yet been investigated systematically. Based on a suite of smoothed particle hydrodynamics simulations with the \swift{} code and a Bondi-Hoyle-Lyttleton subgrid gas accretion model, we investigate the impact of repositioning on SMBH growth and on other baryonic components through AGN feedback. Across at least a factor $\sim$1000 in mass resolution, SMBH repositioning (or an equivalent approach) is a necessary prerequisite for AGN feedback; without it, black hole growth is negligible. Limiting the effective repositioning speed to $\lesssim$10 km s$^{-1}$ delays the onset of AGN feedback and severely limits its impact on stellar mass growth in the centre of massive galaxies. Repositioning has three direct physical consequences. It promotes SMBH mergers and thus accelerates their initial growth. In addition, it raises the peak density of the ambient gas and reduces the SMBH velocity relative to it, giving a combined boost to the accretion rate that can reach many orders of magnitude. Our results suggest that a more sophisticated and/or better calibrated treatment of SMBH repositioning is a critical step towards more predictive galaxy formation simulations.
\end{abstract}




\section{Introduction}
Energy feedback from active galactic nuclei (AGN) that are powered by supermassive black holes (SMBHs) at the centres of massive galaxies is believed to be a key driver of galaxy formation and evolution. On the observational side, evidence in support of this mechanism includes the tight relation between SMBH mass and the stellar mass or velocity dispersion of their host galaxy (e.g.~\citealt{Magorrian_et_al_1998,McConnell_Ma_2013, Marasco_et_al_2021}); the presence of high-velocity gas outflows in galaxies hosting AGN (e.g.~\citealt{Genzel_et_al_2014}); and the ubiquity of low-density X-ray cavities in galaxy clusters combined with the typically low star formation rate of their central galaxies (e.g.~\citealt{Fabian_et_al_2006,McNamara_Nulsen_2007,Hlavacek-Larrondo_et_al_2012}). On the theoretical side, both semi-analytic models and hydrodynamic simulations have found that the inclusion of AGN feedback is essential to reproduce observations of e.g.~the stellar masses, star formation rates, and X-ray properties of massive galaxies and their host haloes (e.g.~\citealt{Bower_et_al_2006,Booth_Schaye_2009,McCarthy_et_al_2010,Crain_et_al_2015,Dubois_et_al_2015,Barai_et_al_2018,Dave_et_al_2019,Zinger_et_al_2020}).

In consequence, almost all contemporary simulations of massive haloes and/or large cosmological volumes contain prescriptions for the formation of, and feedback from, SMBHs (for a recent review see \citealt{Vogelsberger_et_al_2020}). Because the characteristic sizes of SMBHs are orders of magnitude smaller than the resolution typical of current large-scale simulations such as \textsc{HorizonAGN} \citep{Dubois_et_al_2014}, \eagle{} \citep{Schaye_et_al_2015}, \textsc{Magneticum} (Dolag et al., in preparation), \textsc{IllustrisTNG} \citep{Pillepich_et_al_2018}, or \textsc{simba} \citep{Dave_et_al_2019}, the seeding of, gas accretion onto, mergers between, and energy feedback from them must all be predicted from resolved scales by means of analytic sub-grid models. The development and improvement of these models is an active area of research (e.g.~\citealt{Booth_Schaye_2009,Rosas-Guevara_et_al_2015,Angles-Alcazar_et_al_2017,
Weinberger_et_al_2017,Henden_et_al_2018,Angles-Alcazar_et_al_2020,Valentini_et_al_2020,Bhowmick_et_al_2021}), with significant differences between the effect of AGN feedback in different simulations (e.g.~\citealt{Davies_et_al_2020,Habouzit_et_al_2021,Oppenheimer_et_al_2021}; see also \citealt{Newton_Kay_2013} and \citealt{Wurster_Thacker_2013} for a comparison of older models).

While this situation is broadly similar to the analogous problem of modelling feedback from star formation, there is an additional subtlety unique to AGN. Because SMBHs represent extreme local mass concentrations, any motion relative to the matter around them leads to momentum exchange via dynamical friction; this slows down the SMBH and hence tends to make it sink towards the local potential minimum. In a collisionless background such as stars or dark matter, dynamical friction is caused by the integrated effect of gravitational two-body encounters between the SMBH and background particles, in analogy to the orbital decay of satellite subhaloes (see e.g.~\citealt{Taffoni_et_al_2003}). For SMBHs moving through a gaseous background, on the other hand, the hydrodynamic disturbance generated by their passage generates a gaseous wake, which similarly leads to a deceleration (e.g.~\citealt{Ostriker_1999,Beckmann_et_al_2018,Morton_et_al_2021}), at least in the absence of feedback (as pointed out by e.g.~\citealt{Gruzinov_et_al_2020} and \citealt{Toyouchi_et_al_2020}).

Neither of these two mechanisms emerges self-consistently from contemporary simulations, because the background cannot be sufficiently well resolved. Individual particles in such a simulation represent a large number of real stars or of any putative dark matter particles, so that the mass contrast between SMBH and background particles is artificially lowered. Together with the need to soften gravitational interactions to avoid spurious two-body relaxation, this suppresses the emergence of dynamical friction effects on SMBHs. Off-centre formation, galaxy mergers, and artificial momentum transfer from other simulated particle species would therefore lead to simulated SMBHs wandering away from the centres of galaxies to a much greater extent than observed (e.g.~\citealt{Reines_et_al_2020, Pesce_et_al_2021}), or than would be expected from physical -- and typically unresolved -- processes such as encounters with dense molecular clouds or gravitational recoil during SMBH mergers (e.g.~\citealt{Campanelli_et_al_2007,Gonzalez_et_al_2007}).

For high-resolution simulations (particle masses $\lesssim\,$10$^4\,\msun$), several groups have recently developed models that attempt to predict and incorporate this unresolved dynamical friction component (e.g.~\citealt{Tremmel_et_al_2015,Pfister_et_al_2019,Dubois_et_al_2021}). These approaches differ in their implementation details as well as in which physical processes they account for: \citet{Tremmel_et_al_2015}, for instance, only consider gravitational dynamical friction and neglect the gas-dynamical effect, whereas \citet{Dubois_et_al_2021} only account for the latter. To our knowledge, none of them (explicitly) consider the role of unresolved matter that is gravitationally bound to the SMBH, such as a nuclear star cluster or gas cloud, which enhances its effective gravitational mass (see e.g.~\citealt{Biernacki_et_al_2017}).

In simulations of representative volumes or zoom-in simulations of galaxy clusters, however, the gap between resolved scales and those down to which dynamical friction originates from is too large and numerical convergence becomes a limiting factor \citep{Tremmel_et_al_2015}. Instead, most of these simulations implicitly assume that dynamical friction is sufficiently efficient to keep SMBHs permanently near the galaxy centre; this is then enforced by periodically or continuously ``repositioning'' SMBH particles towards their local potential minimum (e.g.~\citealt{Springel_et_al_2005,Booth_Schaye_2009,Vogelsberger_et_al_2013,Rasia_et_al_2015,Schaye_et_al_2015,Weinberger_et_al_2017,Dave_et_al_2019,Bassini_et_al_2020}).

Most commonly, SMBHs are repositioned to the neighbouring resolution element with the lowest gravitational potential, as done in e.g.~the simulations of \citet{Springel_et_al_2005},  \citet{Booth_Schaye_2009}, \eagle{} \citep{Schaye_et_al_2015}, \bahamas{} \citep{McCarthy_et_al_2017}, \textsc{IllustrisTNG} \citep{Weinberger_et_al_2017}, and \textsc{Fable} \citep{Henden_et_al_2018}. In the latter, the SMBH velocity is additionally changed to the local mass-weighted average. In a slightly different manner, SMBHs in \textsc{Simba} \citep{Dave_et_al_2019} are repositioned to the potential minimum of their FOF host group (provided it is not too far away), and their velocity set to the FOF centre-of-mass velocity. SMBH repositioning is also used in several higher-resolution zoom-in suites, such as \textsc{Apostle} \citep{Sawala_et_al_2016}, \textsc{Auriga} \citep{Grand_et_al_2017}, and \textsc{Artemis} \citep{Font_et_al_2020}. Two notable exceptions using neither repositioning nor explicit forms of dynamical friction modelling for SMBHs are the \textsc{HorizonAGN} \citep{Dubois_et_al_2012,Dubois_et_al_2014} and \textsc{magneticum} \citep[Dolag, priv. comm.]{Steinborn_et_al_2015} simulation suites. \textsc{Magneticum} instead attempts to avoid SMBHs wandering off from the centre of their host galaxy in the first place, by seeding them at the location of the star particle with the lowest gravitational potential and avoiding strong artificial kicks during e.g.~gas accretion and mergers (for details, see \citealt{Hirschmann_et_al_2014}).

Despite its common use, the assumption of instantaneous repositioning is not justified by analytic expectations, since a typical SMBH orbit should only decay on a Gyr time scale \citep{Tremmel_et_al_2015,Pfister_et_al_2019}, at least when effects other than gravitational dynamical friction from the SMBH itself are neglected. \citet{Ma_et_al_2021} found that the situation is even worse for newly formed SMBH ``seeds'', whose low mass should give rise to such weak dynamical friction that they are not expected to sink to the galaxy centre at all. Several recent works have highlighted the severe impacts of this overly simplistic and efficient SMBH repositioning on simulation predictions for the SMBH population themselves, such as their merger rates (e.g.~\citealt{Volonteri_et_al_2020}) or the fraction of galaxies with off-centre AGN (e.g.~\citealt{Bellovary_et_al_2019,Boldrini_et_al_2020,Bartlett_et_al_2021,Bellovary_et_al_2021,Ricarte_et_al_2021}). However, as discussed at the beginning, SMBHs are included in large-scale simulations not only -- and arguably not even primarily -- as objects of interest in their own right, but due to their fundamental importance for massive galaxies and their host haloes. It is conceivable that AGN feedback itself is also sensitive to the treatment of unresolved SMBH dynamics, for example due to the dependence of SMBH accretion rates on the ambient gas properties. To our knowledge, however, no detailed exploration of this dependence is reported in the literature so far, in contrast to other aspects of the SMBH model such as gas accretion (e.g.~\citealt{Booth_Schaye_2009,Rosas-Guevara_et_al_2015,Angles-Alcazar_et_al_2017}) or feedback (e.g.~\citealt{Sijacki_et_al_2007,Crain_et_al_2015,Weinberger_et_al_2018,Su_et_al_2021}).

The objective of this paper is therefore to test the impact of a simplified SMBH dynamics modelling approach via simple repositioning on other components of the simulation. We could attempt to achieve this by comparing to a ``gold-standard'' simulation with detailed, explicit modelling of dynamical friction on SMBHs. However, the required high resolution of such a run would not only make it computationally prohibitive, but also complicate comparisons to the lower-resolution simulations of interest for modelling large cosmological volumes: since galaxy formation models are typically not numerically converged (see e.g.~the discussion in section 2.2 of \citealt{Schaye_et_al_2015}), the resolution difference would inevitably introduce additional differences that are difficult to distinguish from the effect of the dynamical friction model. Instead, we will base our investigation on comparisons between simulations at fixed (intermediate and low) resolution that differ only in their modelling of SMBHs. While this approach does not permit us to identify an ``ideal'' simplified way of modelling SMBH dynamics (if one even exists), it does provide a direct quantitative assessment of the difference between these prescriptions and hence the systematic uncertainty associated with their inclusion in contemporary large-scale simulations.

The remainder of this paper is structured as follows. In Section \ref{sec:model}, we describe the simulation model and setup that we use, and then present our comparison of different SMBH repositioning schemes in Section \ref{sec:comparison}. We analyse the physical mechanisms by which repositioning affects the simulation in Section \ref{sec:analysis}, before summarizing our results in Section \ref{sec:summary}. In an appendix, we present a comparison between the SMBH model employed here and that used in the \eagle{} simulations.

\section{Simulation model and initial conditions}
\label{sec:model}
Our simulations have been performed with \swift{} \citep{Schaller_et_al_2018}, a novel $N$-body gravity and smoothed particle hydrodynamics (SPH) solver  built around a fine-grained tasking framework with a hybrid parallelisation approach (see \citealt{Schaller_et_al_2016} for a description of an earlier version)\footnote{\swift{} is publicly available at \url{http://www.swiftsim.com}, including the full subgrid physics model used here.}. The code not only achieves excellent weak and strong scaling \citep{Borrow_et_al_2018}, but is also highly modular with a number of implemented gravity, hydrodynamics, and subgrid physics schemes. Here, we use the Fast Multipole Method \citep{Greengard_Rokhlin_1987} for gravity, fully adaptive time steps with the \citet{Durier_DallaVecchia_2012} limiter, and the \sphenix{} SPH scheme \citep{Borrow_et_al_2020} with the \citet{Wendland_1995} $C^2$ kernel and a target smoothing length of 1.2348 times the local inter-particle separation -- corresponding to an average approximate neighbour number of 65 for a uniform field -- as hydrodynamics solver. In the remainder of this section, we first briefly summarize our choice of subgrid models for processes not related to SMBHs (Sec.~\ref{sec:sim_subgrid}), then provide a detailed description of our SMBH model (Sec.~\ref{sec:sim_bh}), and finally describe the initial conditions from which our simulations are evolved (Sec.~\ref{sec:sim_ics}).

\subsection{Simulation model}
\label{sec:sim_subgrid}
Our suite of subgrid physics models is based on those used in the \textsc{owls} \citep{Schaye_et_al_2010} and \eagle{} \citep{Schaye_et_al_2015} simulations, with a number of mostly minor modifications. Apart from the SMBH model that we describe in detail below, only a succinct summary is provided here; the interested reader is referred to their detailed description and justification in \citet[see also \citealt{Crain_et_al_2015}]{Schaye_et_al_2015}.

Radiative cooling and photoheating are implemented on an element-by-element basis using pre-computed tables based on \textsc{cloudy} that also account for a time-dependent UV/X-ray background, inter-stellar radiation field, and self-shielding of dense gas \citep{Ploeckinger_Schaye_2020}. In contrast to the \citet{Wiersma_et_al_2009a} tables used in \eagle{}, they can model gas cooling down to a temperature of 10 K. The resolution of our simulations is, however, not sufficient to resolve the cold dense phase of the inter-stellar medium; we therefore follow \citet{Schaye_DallaVecchia_2008} and use an effective pressure floor (implemented as an entropy floor) with a slope of $\gamma = 4/3$ at densities $n_\mathrm{H} \geq 10^{-4} \mathrm{cm}^{-3}$, normalised\footnote{The slope and normalisation of this pressure floor are identical to those of the \eagle{} Reference model, but the threshold density is lower by a factor of $10^3$. This is because we no longer employ a second floor at $T = 8000$ K that was necessary in \eagle{} to prevent the formation of a tenuous cold gas phase. In practice, this change has no impact on our results.} to $T = 8000$ K at a density $n_\mathrm{H} = 0.1 \mathrm{cm}^{-3}$. Following the recent measurements by the \textit{Planck} satellite, hydrogen reionization is assumed to occur at redshift $z = 7.5$ \citep{Planck_2018}.

As in \eagle{}, star formation is implemented stochastically with the \citet{Schaye_DallaVecchia_2008} pressure-law implementation of the \citet{Kennicutt_1998} relation between star formation rate and gas surface density. Star forming gas particles are identified based on the properties of their sub-grid cold ISM phase as predicted by the \citet{Ploeckinger_Schaye_2020} cooling tables, with the sub-grid cold ISM temperature $T_\mathrm{subgrid}$ taken as that corresponding to pressure and thermal equilibrium (we select the lowest one if more than one solution exists; the corresponding equilibrium density is identified as the sub-grid density). Gas particles are star forming if they are within 0.3 dex from the entropy floor and either their sub-grid temperature $T_\mathrm{subgrid} < 1000$ K or their subgrid density $n_\mathrm{H} > 10\, \mathrm{cm}^{-3}$ and simultaneously $T_\mathrm{subgrid} < 10^{4.5}$ K. In practice, this selection is similar to the  metallicity-dependent density threshold of \citet{Schaye_2004} that was used in \eagle{}.

Energy feedback from star formation is implemented in stochastic thermal form \citep{DallaVecchia_Schaye_2012}, with stars of initial masses between 8 and 100 $\msun$ assumed to explode as core-collapse supernovae (SNe). We use the same scaling law as for \eagle{} to adjust the SN energy to the birth density and metallicity of each star particle \citep{Crain_et_al_2015}, but with the energy efficiency scaled between 0.5 and 5.0 (instead of 0.3--3.0) to compensate for the higher minimum SNe progenitor mass (8 instead of 6 $\msun$). Two further changes compared to \eagle{} are that we sample the delay between star formation and feedback according to the expected life times of the SNe progenitors, and that the feedback energy is injected into the nearest gas neighbour, rather than a randomly chosen particle inside the smoothing kernel of the star particle; the latter change has a non-trivial impact as investigated in detail by \citet{Chaikin_et_al_2022}. We also model energy feedback from type Ia SNe and AGB stars, in addition to mass and metal injection into their surroundings based on the prescription of \citet{Wiersma_et_al_2009b}. Since the latter can increase the mass of individual gas particles near the centre of massive galaxies by large factors ($\gtrsim 10^3$), gas particles are split in two when their mass exceeds 4 times the initial baryon mass.

\subsection{Modelling of supermassive black holes}
\label{sec:sim_bh}

Like the remainder of our subgrid model suite, our treatment of SMBHs is based on the \eagle{} model as described by \citet{Schaye_et_al_2015}, which is itself based on \citet{Booth_Schaye_2009} and earlier work by \citet{Springel_et_al_2005}. Because of the key relevance of these prescriptions to our work, and a number of changes that we have made compared to \eagle{}, we now describe them in detail; they cover the seeding of SMBHs, their repositioning, mergers between them, growth by gas accretion, and the injection of feedback into their surrounding gas.

\subsubsection{SMBH seeding}
The initial formation of SMBH seeds is an unsolved problem and several mechanisms are considered as potentially viable (e.g.~\citealt{Latif_Ferrara_2016}). These include remnants of Population III stars \citep{Smith_et_al_2018}, direct collapse due to cooling in metal-free gas haloes \citep{Bromm_Loeb_2003}, and runaway collapse of nuclear star clusters as a consequence of gas accretion \citep{Davies_et_al_2011}. Common to all of them is that they cannot be modelled self-consistently at the resolution achieved by current large cosmological volumes (see \citealt{Tremmel_et_al_2017} for an alternative applicable to higher-resolution simulations). Since observed SMBHs clearly must have formed in some way, we therefore follow the well-established practice of injecting SMBH seeds by hand into sufficiently massive haloes \citep{Springel_et_al_2005}.

For this purpose, we periodically run an on-the-fly friends-of-friends (FOF) group finder (see \citealt{Willis_et_al_2020} for the implementation in \swift{}) with a linking length of 0.2 times the mean inter-particle separation, to identify haloes with a total mass above a threshold $M_\mathrm{FOF}$ (see Table \ref{tab:parameters}). In each such halo that does not already contain at least one SMBH particle, we then convert the densest gas particle into an SMBH, which retains the mass and phase-space coordinates of its parent gas particle.

The mass of the newly formed SMBH particle is, however, not characteristic of (putative) real SMBH seeds; it is also explicitly resolution dependent and potentially already close to the $\sim\!\!10^7\,\msun$ scale characteristic for galaxies in which AGN feedback is important (e.g.~\citealt{McConnell_Ma_2013, Sahu_et_al_2019}). Instead of directly associating the particle (dynamical) mass to the SMBH itself, we therefore assign a (lower) subgrid value $m_\mathrm{seed}$ to the latter, which is chosen to be compatible with at least some theoretical seed formation scenarios and well below the mass scale at which AGN feedback becomes important. SMBH seeds can then be placed in appropriately low-mass haloes and grow (together with their host galaxy) until the point at which their feedback becomes important; the difference between dynamical and subgrid mass is assumed to represent a subgrid gas reservoir around the SMBH (see below).

The choices of the halo mass threshold for SMBH seeding $M_\mathrm{FOF}$ and the seed mass $m_\mathrm{seed}$ have a significant impact on SMBHs in low-mass galaxies (e.g.~\citealt{Booth_Schaye_2009}), although not on the onset of rapid SMBH growth \citep{Booth_Schaye_2009,Bower_et_al_2017}. We therefore calibrate these values roughly to achieve an approximately realistic SMBH mass ($\lesssim\,10^5\,\msun$) below a galaxy stellar mass of $\mstar \sim 10^{10}\,\msun$ in our baseline model (see Table \ref{tab:parameters}); the \eagle{} reference model used $m_\mathrm{seed} = 1.47 \times 10^5\,\msun$.
   
\subsubsection{Repositioning to account for unresolved dynamical friction}
Simulations at the resolution that we consider here can neither model dynamical friction on SMBHs self-consistently, nor can they reliably predict its effect in a sub-grid fashion (e.g.~\citealt{Tremmel_et_al_2015}). Instead, we assume that the net effect of dynamical friction is to move them towards the local potential minimum and directly impose repositioning in a simple way. First, we identify all gas and (other) SMBH particles that lie within the kernel support radius of the SMBH under consideration\footnote{The smoothing kernel for SMBHs uses the same \citet{Wendland_1995} $C^2$ functional form and effective number of neighbours as for the hydrodynamics. We point out that neighbours are defined not in terms of the smoothing length $h$, but the radius $H$ within which the kernel function is non-zero (see section 2.1 of \citealt{Dehnen_Aly_2012}). We also note that, unlike e.g.~\textsc{simba} \citep{Dave_et_al_2019}, we do not impose a maximum on the SMBH smoothing length other than a technical limit of 0.5 comoving Mpc.} and that are within at most three gravitational softening lengths $\epsilon_\mathrm{baryon}$. Out of these, we then find the particle with the lowest gravitational potential $\Phi_\mathrm{min}$. Our baseline model (``\texttt{Default}'') is then to move the SMBH immediately to the coordinates of this particle, provided that $\Phi_\mathrm{min} < \Phi_\mathrm{SMBH}$.

This approach is similar to what was done in \eagle{} and most other contemporary simulations, as discussed in the Introduction. However, there are three subtle differences of our repositioning scheme compared to \eagle{}. Firstly, the \eagle{} model also allowed repositioning towards neighbouring star and dark matter particles. The reason for changing this here is purely technical; \swift{} searches for neighbours separately by type and only gas and SMBH neighbours are required for other parts of the model (see below). We have, however, performed extensive tests with a modified code version that considers all particle types for repositioning and found essentially no difference; in other words, SMBHs typically have enough nearby gas particles that these allow an efficient migration towards the potential minimum. 

Secondly, in contrast to \eagle{} we apply repositioning to all SMBHs, and not only to those with a mass $m_\mathrm{BH} < 100\, m_\mathrm{gas}$. This is because even those very massive SMBHs are subject to unresolved dynamical friction, be it from additional (gas and stellar) mass bound to them, from gas-dynamical effects \citep{Ostriker_1999}, or from remaining softening of gravitational interactions with its neighbours. For the simulations presented here, this change is of little relevance: hardly any SMBHs reach such high masses.

Finally, repositioning in \eagle{} -- and also in many other simulations\footnote{Simulations that do \emph{not} apply such a restriction when repositioning SMBHs include those of \citet[V.~Springel, priv.~comm.]{Springel_et_al_2005}, \textsc{IllustrisTNG} \citep{Weinberger_et_al_2017}, \textsc{Auriga} (R. Grand, priv. comm.), and \textsc{Simba} \citep{Dave_et_al_2019}.} such as those of \citet{Booth_Schaye_2009}, \bahamas{} \citep{McCarthy_et_al_2017}, and \textsc{Artemis} \citep{Font_et_al_2020} was limited to particles moving with a speed less than $0.25\,c_\mathrm{sound}$ with respect to the SMBH, where $c_\mathrm{sound}$ is the sound speed of its surrounding gas (see below). This speed threshold was motivated by the desire to prevent unphysical SMBH jumps between galaxies during close flybys. We have found this worry to be largely unjustified: when such jumps occur, the fact that the velocity of the SMBH is not \emph{directly} affected by the repositioning in our approach\footnote{Effectively, we have found that SMBHs come to rest in the frame of their host halo within $\lesssim\!\!10$ Myr once they have reached its potential minimum: their initial kinetic energy is rapidly lost as they attempt to climb out of the potential well, only to be continuously repositioned back to its deepest point.} means that it will typically move back towards its original galaxy within a few time steps. Only for low-mass satellite galaxies is the effect more significant: their SMBHs sometimes transfer prematurely to the central galaxy during close pericentric passages. A few per cent of galaxies with $\mstar \lesssim 10^{10}\,\msun$ are therefore left without a SMBH (see Appendix \ref{app:bh_free_galaxies}). In our baseline model we therefore allow repositioning irrespective of the velocity of the target particle. To explore the effects of this speed threshold, we also consider two variants in which repositioning is limited to particles moving at speeds less than $0.25$ and $0.5\,c_\mathrm{sound}$ with respect to the SMBH, respectively (``\texttt{ThresholdSpeed0p25cs}'' and ``\texttt{ThresholdSpeed0p5cs}'').

In addition to these two variants, we include models with three different repositioning approaches. The first, ``\texttt{NoRepositioning}'' does not apply any explicit change to SMBH positions and relies purely on the resolved dynamics of the simulation to predict their motion. This provides a comparison against which to evaluate the impact of other approaches.

Secondly, we consider an approach where the SMBH is \emph{not immediately} moved to the position of its lowest-potential neighbour; instead, we compute the distance travelled at a putative drift speed $v_\mathrm{drift}$ over one time step $\Delta t$, and then move it by (at most) this distance (or less, if it is already closer to it than $v_\mathrm{drift}\, \Delta t$). Based on the analytic sinking time scale of \citet{Taffoni_et_al_2003}, \citet{Tremmel_et_al_2015}, and \citet{Pfister_et_al_2019}, a SMBH of  $m_\mathrm{BH} = 10^6\,\msun$ in a Milky Way like galaxy has -- neglecting gas-dynamical effects and additional mass bound to the SMBH -- a sinking time scale that corresponds to $v_\mathrm{drift} \sim 10\, \mathrm{km}\, \mathrm{s}^{-1}$. Bracketing this fiducial value, we test five choices of $v_\mathrm{drift} = \{2, 5, 10, 50, 250\}\,\mathrm{km}\, \mathrm{s}^{-1}$ and denote these models as ``\texttt{DriftSpeed2kms}'', ``\texttt{DriftSpeed5kms}'', ``\texttt{DriftSpeed10kms}'', ``\texttt{DriftSpeed50kms}'', and ``\texttt{DriftSpeed250kms}'', respectively.

It may be tempting to view these models as a step towards a more realistic modelling of SMBH dynamics also in relatively low resolution simulations. We emphasize, however, that this approach is nonetheless not particularly realistic: for example, while we impose a fixed drift speed for all SMBHs, dynamical friction should generally be stronger, and $v_\mathrm{drift}$ therefore higher, for more massive SMBHs and those in denser environments (the quantitative form of these dependencies is, however, less clear). Our aim here is to provide a first-order test of the sensitivity of simulation predictions to a more gradual repositioning; we therefore defer tests of more complex models to future work.   

Finally, we consider a variant in which repositioning is limited to SMBHs that have not yet grown significantly, i.e.~$m_\mathrm{BH} < 1.2\,m_\mathrm{seed}$ (``\texttt{SeedRepositioningOnly}''). This will allow us to differentiate between the initial repositioning of SMBH seeds from their formation site to the centre of their halo and the continuous ``pinning'' of evolved SMBHs to this position.

We note that, in all simulations presented here, the potential $\Phi$ of neighbouring particles that determines where to reposition the SMBH to also includes the contribution from the SMBH itself. To our knowledge this is also the case for other simulations in which repositioning is based on the local potential minimum, but it is not actually desirable: sufficiently massive SMBHs can create a local potential minimum and therefore be prevented from repositioning towards the actual centre of their host halo. We have verified that the better approach of subtracting the contribution of the SMBH to the neighbour potential when selecting the lowest potential neighbour does not affect the results shown here. It does, however, lead to appreciably lower star formation rates of massive haloes ($M \gtrsim 10^{14}\,\msun$) in low-resolution simulations, due to more efficient repositioning of their SMBH towards the halo centre -- at least as long as the resolution is not so low that even the most massive black holes never grow significantly above the mass of gas particles. For new simulations we recommend subtracting the contribution of the SMBH to the gravitational potential for the purpose of repositioning.

\subsubsection{Mergers between SMBHs}
Through mergers, massive haloes can contain multiple SMBHs. Under favourable circumstances, two or more of these can become gravitationally bound to each other; if they can continue losing orbital energy through e.g.~dynamical friction or three-body encounters with stars or gas clouds, they may eventually approach each other closely enough to emit gravitational waves and merge (e.g.~\citealt{PortegiesZwart_McMillan_2000,LIGO_Virgo_2016}). The resolution of our simulations permits only a severely oversimplified modelling of this process: once a SMBH ($A$) comes within three baryonic softening lengths of a more massive one ($B$) and is also within $B$'s kernel support radius $H_B$, the two are merged instantaneously if, and only if, their relative velocity $\Delta v_{AB}$ is below the escape velocity from the more massive SMBH at their current distance $r_{AB}$: $\Delta v^2_{AB} < 2Gm_\mathrm{BH,\,B} / r_{AB}$; i.e. if they are gravitationally bound rather than just a fly-by. Although this criterion actually better corresponds to the (slightly less restrictive) velocity threshold $\Delta v^2_{AB} < 2G(m_\mathrm{BH,\,A} + m_\mathrm{BH,\,B}) / r_{AB}$, we have verified that our results are insensitive to this difference.

This criterion is different in detail from \eagle{}, where SMBH mergers required a velocity difference smaller than the circular velocity at the edge of $B$'s smoothing kernel. Especially for SMBHs with a low ambient gas density, the kernel radius can be more than an order of magnitude larger than the gravitational softening length, so that our approach allows a larger fraction of SMBH pairs to merge (or rather, it typically allows them to merge more quickly). In a merger, we transfer the momentum, subgrid mass, and particle mass of the lower-mass SMBH ($A$) to the more massive one ($B$), and then remove it from the simulation; this makes it straightforward to track the evolution of SMBH main progenitors back in time. 

In rare situations, more than two SMBHs can be eligible to merge with each other in the same time step; there are numerous possible configurations depending on how many there are and which of them are mutually eligible to merge with each other. In such situations, each SMBH is only swallowed once, by the most massive SMBH eligible for it, and a SMBH that is due to be swallowed by one that is simultaneously swallowed itself is left intact, at least until the next time step.

\subsubsection{Gas accretion onto SMBHs}
Apart from mergers, SMBHs grow continuously by accreting gas from their surroundings. The resolution of our simulations only permits a crude estimate of this accretion rate $\dot{m}_\mathrm{BH}$ based on the spherically symmetric Bondi-Hoyle-Lyttleton model \citep{Hoyle_Lyttleton_1939,Bondi_Hoyle_1944}, limited to the \citet{Eddington_1926} rate,
\begin{equation}
	\dot{m}_\mathrm{BH} = \mathrm{min}\left[\alpha \cdot \frac{4\mathrm{\uppi} G^2 \cdot m_\text{BH}^2 \cdot \rho_\mathrm{gas}}{\left(c_\mathrm{sound}^2 + v_\text{gas}^2 \right)^{3/2}}, \,\dot{m}_\mathrm{Edd}\right],
	\label{eq:bondi}
\end{equation}
where $G$ is Newton's constant, $\rho_\mathrm{gas}$ and $c_\mathrm{sound}$ are the density and sound speed of the ambient gas, respectively, and $v_\mathrm{gas}$ is the bulk velocity of the gas relative to the SMBH. The latter two are computed as kernel-weighted averages over all gas neighbours while $\rho_\mathrm{gas}$ is computed in analogy to the density computation of gas particles by the hydrodynamics solver\footnote{In other words, we do not average the densities of individual neighbour particles, but estimate the gas density at the SMBH position -- which is the relevant quantity in the Bondi-Hoyle-Lyttleton model -- in the standard SPH way.}. For gas particles within 0.3 dex of the entropy floor the temperature, and hence sound speed, is artificially raised, which could strongly suppress the estimate of $\dot{m}_\mathrm{BH}$; we therefore assume a sound speed corresponding to $T = 8000$ K for these particles, as appropriate for the warm volume-filling phase of the inter-stellar medium.  

The limiting Eddington rate is calculated as
\begin{equation}
	\dot{m}_\mathrm{Edd} = \frac{4 \uppi G m_\mathrm{p}}{\epsilon_\mathrm{r} c \sigma_\mathrm{T}} m_\mathrm{BH}, 
\end{equation}
with proton mass $m_\mathrm{p}$, radiative efficiency $\epsilon_\mathrm{r} = 0.1$, speed of light $c$ and Thomson cross-section $\sigma_\mathrm{T}$. Since $\dot{m}_\mathrm{Edd} \propto m_\mathrm{BH}$, Eddington-limited accretion leads to exponential SMBH growth with an e-folding time of $(\epsilon_\mathrm{r} c \sigma_\mathrm{T})/(4\pi G m_\mathrm{p}) = 45\, \mathrm{Myr}$ (the Salpeter time).

The boost factor $\alpha$ in equation \ref{eq:bondi} is commonly used to compensate for the artificially low gas densities near the centre of galaxies in low-resolution simulations. Following \citet{Booth_Schaye_2009}, we make $\alpha$ dependent on the gas density,
\begin{equation}
	\alpha = \mathrm{max}\left[(n_\mathrm{H}/n_\mathrm{H}^\star)^\beta, 1\right]
	\label{eq:bh_boost}
\end{equation}
where $n_\mathrm{H}$ is the ambient number density of hydrogen atoms, $n_\mathrm{H}^\star = 0.1\,\mathrm{cm}^{-3}$ is a reference density above which an (unresolved) cold interstellar gas phase is expected to form \citep{Schaye_2004}, and $\beta$ is a free parameter. No boost is therefore applied for SMBHs in (well-resolved) low-density environments. With $\beta = 0$, $\alpha = 1$ everywhere, i.e.~the unboosted Bondi-Hoyle-Lyttleton accretion rate is used, as was the case in \eagle{} \citep{Schaye_et_al_2015}. We use the same setting for our intermediate-resolution simulations, but have found $\beta > 0$ necessary for  low-resolution runs (see Table \ref{tab:parameters}).

Different from \eagle{}, our SMBH accretion model does not include a suppression factor related to the angular momentum of gas around the SMBH \citep{Rosas-Guevara_et_al_2015}. There are two reasons for this change. Firstly, it has become increasingly clear that the resolved gas kinematics around SMBHs (typically on kpc scales) are only poorly correlated with the pc scale structure of the actual accretion disc (e.g.~\citealt{Angles-Alcazar_et_al_2020}). We therefore cannot reliably predict the angular momentum of, and hence efficiency of viscous gas transport through, the subgrid accretion disc on which the \citet{Rosas-Guevara_et_al_2015} model is based (even the Bondi radius of all but the most massive SMBHs is typically well below the resolution scale, see e.g.~eq. (1) of \citealt{Rosas-Guevara_et_al_2015}). Secondly, \citet[see also \citealt{Dubois_et_al_2015}]{Bower_et_al_2017} have shown that the onset of AGN feedback, which this suppression factor was intended to regulate, is instead more fundamentally determined by the decreasing efficiency of SN feedback; whenever the latter fails to eject gas from the halo, gas densities increase and SMBHs grow rapidly, until their own feedback is able to regulate the gas content. Accordingly, we have found that calibrating the SN feedback scaling alone, in addition to the SMBH seeding parameters, provides good control of AGN feedback on intermediate mass scales. In Appendix \ref{app:bh_changes} we discuss the effects of this change, as well as the other main deviations in our SMBH model from \eagle{}, in more detail.

Out of the total gas mass accreted over a time step $\Delta t$, $\Delta m = \dot{m}_\mathrm{BH}\,\Delta t$, a fraction $\epsilon_\mathrm{r}$ is converted to energy (see below). The remainder, $(1-\epsilon_\mathrm{r})\,\Delta m$ is added to the subgrid mass $m_\mathrm{BH}$ of the SMBH. If its particle mass is greater than the new $m_\mathrm{BH}$, the accreted mass is assumed to come from the remaining sub-grid gas reservoir when the SMBH was seeded (see above), i.e.~from within the SMBH particle itself. To account for the fraction $\epsilon_\mathrm{r}$ of the accreted mass that is converted to energy, the dynamical SMBH mass is reduced by $\epsilon_\mathrm{r}\,\Delta m$ (this was not done in \eagle{}).

Otherwise (i.e.~if there is insufficient mass left in the sub-grid gas reservoir), the mass deficit $m_\mathrm{BH} + \Delta m - m_\mathrm{BH}^\mathrm{dynamical}$ (the last term denotes the dynamical mass of the SMBH) is drawn from the surrounding gas particles. In \eagle{}, this was done by stochastically swallowing individual gas neighbours. This is not an ideal approach: the momentum imparted on the SMBH from the swallowed gas particle may artificially dislodge it from its position, particularly without instantaneous repositioning (as also discussed by \citealt{Steinborn_et_al_2015}). The mass of a gas particle is also typically much greater than $\Delta m$, so that the dynamical mass of SMBH particles remains systematically above its subgrid mass. Both issues become more severe when individual gas particles have been enriched to masses well above their initial value due to stellar outflows, which is particularly common in massive, gas-poor galaxies. Instead of swallowing entire particles, we therefore transfer a (typically very) small fraction of mass from all gas neighbours to the SMBH simultaneously, with the mass $\delta m_i$ ``nibbled'' from each neighbour $i$ weighted in analogy to their contribution to the gas density at the location of the SMBH,
\begin{equation}
	\delta m_i = (1-\epsilon_\mathrm{r}) \Delta m \left[\frac{w_i m_i}{\sum_j (w_j m_j)}\right],
\end{equation} 
where $w_i$ is the kernel weight of particle $i$, $m_i$ its mass, and the sum is over all neighbours; as above, the factor of $(1-\epsilon_\mathrm{r})$ accounts for the mass converted to energy. In addition to mass, a fraction $\delta m_i / m_i$ of the momentum of neighbour $i$ is also transferred to the SMBH. To prevent individual gas particles from becoming too light, we exclude any neighbour that would be reduced to less than half its initial mass and accept that the dynamical mass of the SMBH grows slightly less than desired in this case. In practice, we have found that this limit is never reached in the simulations presented here, because stars typically inject far more mass into gas particles than is drained by SMBHs.

\subsubsection{Thermal energy feedback from SMBHs}
To model AGN feedback, we inject energy $\Delta E = \epsilon_\mathrm{r}\epsilon_\mathrm{f}\,\Delta m\, c^2$ into the surrounding gas, where the factor $\epsilon_\mathrm{f}$ accounts for the (uncertain) coupling efficiency; we use $\epsilon_\mathrm{f} = 0.1$ throughout our simulations, but note that changes to this parameter only significantly affect the masses of SMBHs \citep{Booth_Schaye_2009, Booth_Schaye_2010}. To avoid strong numerical cooling losses \citep{DallaVecchia_Schaye_2012}, we follow \citet{Booth_Schaye_2009} and heat particles by a large temperature increment $\Delta T_\mathrm{AGN} = 10^{8.5}$ K, which on average requires an energy $E_1 = \overline{m}_\mathrm{ngb}\, \Delta T_\mathrm{AGN} k_\mathrm{B} / [\mu m_\mathrm{p}\, (\gamma - 1)]$ per particle, where $\overline{m}_\mathrm{ngb}$ is the (unweighted) average mass of the gas particles within the SMBH kernel, $k_\mathrm{B}$ is Boltzmann's constant, $\mu = 0.59$ is the molecular mass of a fully ionized primordial gas,  $m_\mathrm{p}$ the proton mass, and $\gamma = 5/3$ the specific heat capacity of an ideal monatomic gas. Because typically $\Delta E \ll E_1$, we store the energy in a subgrid reservoir $E_\mathrm{BH}$ until $E_\mathrm{BH} \geq N_\mathrm{heat}\, E_1$. In our simulations, we exclusively use $N_\mathrm{heat} = 1$, but higher values have been used in previous simulations (e.g.~$N_\mathrm{heat} = 10$ for \bahamas{}; \citealt{McCarthy_et_al_2017}) to make the feedback more explosive. When this condition is satisfied -- i.e.~when there is enough energy available to heat at least $N_\mathrm{heat}$ average-mass neighbour particles -- we heat the $N = \mathrm{min}\left(\lfloor E_\mathrm{BH} / E_1\rfloor, 50, N_\mathrm{ngb}\right)$ particles\footnote{The maximum value of $N = 50$ is set for technical reasons only; we have verified that it does not influence our results.} by $\Delta T$, where $N_\mathrm{ngb}$ is the total number of neighbours within the SMBH kernel, and subtract the actually used energy from $E_\mathrm{BH}$. To reduce the probability of having to heat more than one particle simultaneously, SMBH time steps are limited such that the expected increase of $E_\mathrm{BH}$ does not exceed\footnote{There is no guarantee that $E_\mathrm{BH} < E_1$, because $\Delta t$ is set based on the current accretion rate $\dot{m}_\mathrm{BH}$, while $\Delta E$ is calculated from $\dot{m}_\mathrm{BH}$ in the next step. To prevent extremely short time steps for massive SMBHs accreting at a high fraction of the Eddington rate, this accretion rate dependent time step floor is not allowed to be $<\! 10^5$ yr.} $E_1$.

In the rare situation where a SMBH grows so rapidly and/or has such a long time step that even heating $\mathrm{min}(N_\mathrm{ngb}, 50)$ average-mass neighbours would require energy $< E_\mathrm{BH}$, we temporarily increase $\Delta T_\mathrm{AGN}$ such that the entire reservoir $E_\mathrm{BH}$ is expected to be used up. This is different from \eagle{}, where only a maximum fraction of 0.3 $N_\mathrm{ngb}$ was targeted for heating in any one time step to avoid energy conservation violations in the pressure-entropy formulation of hydrodynamics used there (see appendix A1.1 of \citealt{Schaye_et_al_2015}). The density-energy based \textsc{sphenix} hydrodynamics scheme that we use is explicitly designed and tested to not require such a limitation \citep{Borrow_et_al_2020,Borrow_et_al_2021}.

The selection of heated particles has two subtle differences compared to \eagle{}. First, there is no stochasticity in how many particles we heat: as in \citet{Booth_Schaye_2009}, this is directly and deterministically set by $E_\mathrm{BH}$ and $E_1$. Second, as for SN feedback, we heat the particle(s) closest to the SMBH rather than selecting at random within the kernel. We test the effect of the latter change in Appendix \ref{app:bh_changes}, but refer the interested reader to Chaikin et al. (in preparation) for a thorough comparison between different thermal feedback injection schemes.

All simulations presented here use the same constant target heating temperature increase of $\Delta T_\mathrm{AGN} = 10^{8.5} \mathrm{K}$. For completeness, we note that our model also includes the option to use an adaptive $\Delta T_\mathrm{AGN}$ that scales with the SMBH mass and ambient density; this is taken as the higher of
\begin{equation}
	\Delta T_\mathrm{vir} = \Delta T_\mathrm{ref}\,\left(\frac{m_\mathrm{BH}}{m_\mathrm{ref}}\right)^{2/3} 
	\label{eq:dtvir}
\end{equation}
and
\begin{equation}
	\Delta T_\mathrm{crit} = 10^{7.5} \mathrm{K}\,\left(\frac{n_\mathrm{H}}{10\, \mathrm{cm}^{-3}}\right)^{2/3} \left(\frac{\overline{m}_\mathrm{ngb}}{10^6\,\msun}\right)^{1/3},
	\label{eq:dtcrit}
\end{equation}
where $\Delta T_\mathrm{ref}$ and $m_\mathrm{ref}$ are free parameters. Equation \eqref{eq:dtvir} is motivated by the scaling relations between SMBH and halo mass, and halo mass and virial temperature, respectively; it aims to keep $\Delta T_\mathrm{AGN}$ well above the halo virial temperature so that feedback-inflated bubbles can buoyantly rise from the halo centre. Equation \eqref{eq:dtcrit} is based on \citet{DallaVecchia_Schaye_2012}, and ensures that the feedback is not rendered numerically inefficient through immediate numerical cooling losses. Finally, $\Delta T_\mathrm{AGN}$ is constrained to lie in the range $[\Delta T_\mathrm{min}, \Delta T_\mathrm{max}]$. With plausible parameters $\{\Delta T_\mathrm{ref}, m_\mathrm{ref}, \Delta T_\mathrm{min}, \Delta T_\mathrm{max}\} = \{10^{8.5}\,\mathrm{K}, 10^8\,\msun, 10^7\,\mathrm{K}, 10^{9.5}\,\mathrm{K}\}$, the simulation outcomes presented here are not noticeably different from our default choice of a constant $\Delta T_\mathrm{AGN} = 10^{8.5}\,\mathrm{K}$, so that we chose the latter for simplicity. It is, however, conceivable that the adaptive $\Delta T_\mathrm{AGN}$ prescription may be advantageous in simulations that include very massive haloes, for which high temperature increases have proven beneficial in terms of their gas content \citep{LeBrun_et_al_2014,Schaye_et_al_2015, Barnes_et_al_2017} and quenched galaxy fractions \citep{Bahe_et_al_2017}; we will explore this in future work.

\subsection{Simulation setup}
\label{sec:sim_ics}

\begin{table*}
	\centering
	\caption{Cosmological, resolution, and subgrid parameters that vary between the two simulation sets used in this work. A prefix of `c' and `p' denotes comoving and proper lengths, respectively.}
	\label{tab:parameters}
	\begin{tabular}{llll} 
		\hline
		Parameter & Meaning & \eagle{}-like & \bahamas{}-like \\
		\hline
		$\Omega_\mathrm{m}$ & Matter density parameter & 0.307 & 0.2793\\
		$\Omega_{\Lambda}$ & Dark energy (DM) density parameter & 0.693 & 0.7207\\
		$\Omega_\mathrm{baryon}$ & Baryon density parameter & 0.04825 & 0.0463 \\
		$h \equiv H_0 / (100\, \mathrm{km}\, \mathrm{s}^{-1} \mathrm{Mpc}^{-1})$ & Reduced Hubble constant & 0.6777 & 0.7\\
		$\sigma_8$ & RMS of the linear matter distribution on a scale $8\,h^{-1}$ cMpc & 0.8288 & 0.821 \\
		$n_\mathrm{s}$ & Spectral index of primordial perturbations & 0.9611 & 0.972\\
		\hline
		$L$ & Box side length & 25 cMpc & 200 cMpc\\
		$N_\mathrm{initial}$ & Initial number of DM and baryon particles (each) & $376^3$ & $360^3$\\
		$m_\mathrm{DM}$ & DM particle mass & $9.77 \times 10^6\,\msun$ & $5.5 \times 10^9\,\msun$\\   
		$m_\mathrm{baryon}$ & Initial baryon particle mass & $1.81 \times 10^6\,\msun$ & $1.1 \times 10^9\,\msun$\\
		$\epsilon_\mathrm{soft,\,DM}$ & Plummer-equivalent gravitational softening length (DM) & min(3.32 ckpc, 1.30 pkpc) & min(22.3 ckpc, 5.7 pkpc) \\
		$\epsilon_\mathrm{soft,\,baryon}$ & Plummer-equivalent gravitational softening length (baryons) & min(1.79 ckpc, 0.70 pkpc) & min(22.3 ckpc, 5.7 pkpc) \\
		\hline
		$M_\mathrm{FOF}$ & Halo mass threshold for SMBH seeding & $10^{10}\,\msun$ & $4\times 10^{11}\,\msun$\\
		$m_\mathrm{seed}$ & SMBH seed mass & $10^4\,\msun$ & $10^5\,\msun$\\
		$\beta_\mathrm{boost}$ & Power-law index of the SMBH accretion boost factor & 0 & 0.6 \\
		\hline 
		
	\end{tabular}
\end{table*}

With the model and its variations described above, we run two sets of periodic-volume simulations. The first uses the \eagle{} L0025N0376 initial conditions as described by \citet{Schaye_et_al_2015}; these fill a cubic volume with a side length of 25 comoving Mpc with (initially) $N = 376^3$ gas particles of mass $m_\mathrm{b} = 1.81 \times 10^6\, \msun$ and an equal number of DM particles with mass $m_\mathrm{DM} = (\Omega_m / \Omega_\mathrm{b} - 1) m_\mathrm{b} = 9.77 \times 10^6\, \msun$. Motivated by \citet{Ludlow_et_al_2019}, who found that the use of more massive DM particles at the same softening length leads to spurious energy transfer from DM to stars, we use a (Plummer-equivalent) softening length for DM particles of 1.3 proper kpc (limited to 3.32 comoving kpc at high redshift) for DM, compared to 0.7 proper kpc (limited to 1.79 comoving kpc) for baryons. We use the same \textit{Planck}-13 cosmology \citep{Planck_2014} as \eagle{} for these simulations; the cosmological parameters are listed in Table \ref{tab:parameters} together with the resolution characteristics and the values for those subgrid parameters that are different between the two simulation sets.

Secondly, we run an equivalent set of simulations at the (much) lower resolution of the large-scale \bahamas{} run \citep{McCarthy_et_al_2017}. These evolve a cubic volume with a side length of 200 comoving Mpc and $N = 360^3$ DM and (initial) gas particles with masses of $5.5 \times 10^9\,\msun$ and $1.1 \times 10^9\,\msun$, respectively. The (Plummer-equivalent) gravitational softening length is 5.7 proper kpc, limited to 22.3 comoving kpc at high redshift, for all particle species. For consistency with the initial conditions, we use the \textit{WMAP}-9 cosmology \citep{Hinshaw_et_al_2013} here, but the small difference from the \textit{Planck}-13 parameters is insignificant for our results. A small number of subgrid parameters relating to AGN feedback have been adjusted compared to the first set to account for the lower resolution; these are listed in the bottom section of Table \ref{tab:parameters}.

For each set, we run twelve simulations from identical initial conditions. Ten of these correspond to the model variations listed above (\texttt{Default, NoRepositioning, ThresholdSpeed0p25cs, ThresholdSpeed0p5cs, DriftSpeed2kms, DriftSpeed5kms, DriftSpeed10kms, DriftSpeed50kms, DriftSpeed250kms, SeedRepositioningOnly}). Two further variations are \texttt{NoAGN}, which is identical to \texttt{Default} except that we disable gas accretion onto, and hence feedback from, SMBHs completely; and \texttt{NoRepos\_MassiveSeeds}, identical to \texttt{NoRepositioning} but with the seed mass increased by $\times 100$. Both will be used below to interpret the results obtained from the other runs.

All simulations are evolved from initial conditions at $z = 127$ to $z = 0$, with five full snapshots stored at $z = \{2, 1, 0.5, 0.2, 0\}$; we use the public \textsc{VELOCIraptor}\footnote{\url{https://velociraptor-stf.readthedocs.io/en/latest/}} structure finder \citep{Elahi_et_al_2019,Canas_et_al_2019} to identify galaxies and haloes in each of these snapshots. A much larger number of outputs store information for only the SMBH particles, whose properties such as ambient density or accretion rate tend to fluctuate rapidly, at 5 Myr intervals.

\section{The impact of black hole repositioning}
\label{sec:comparison}
\subsection{Dependence of galaxy properties on repositioning}
\begin{figure*}
	\includegraphics[width=2.1\columnwidth]{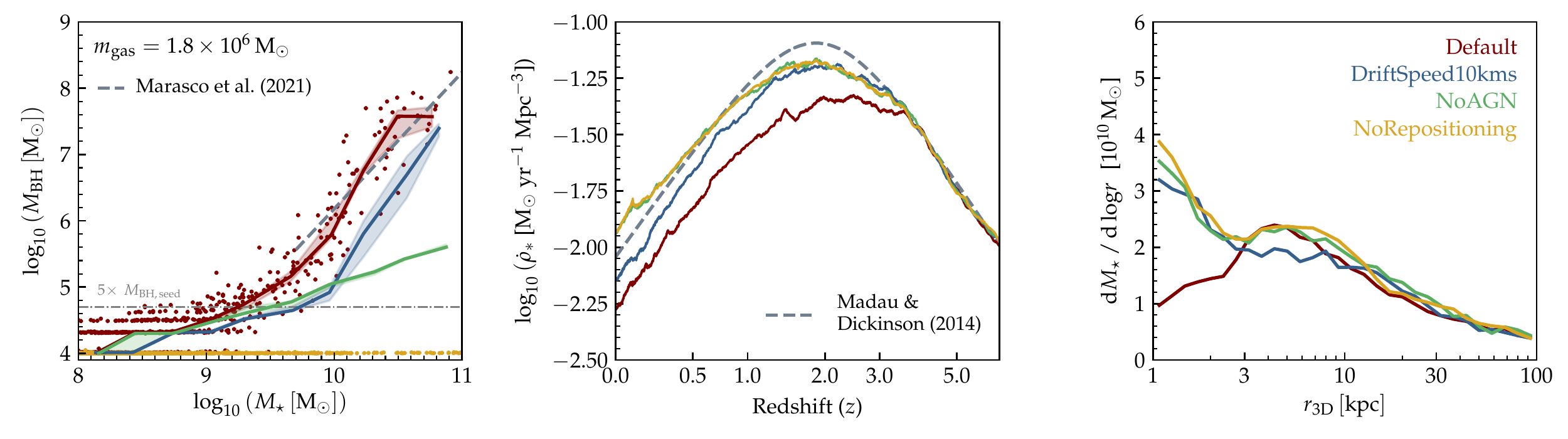}
	\caption{Comparison between different repositioning approaches for \eagle{}-resolution simulations. From left to right, the three panels show the $z = 0$ stellar mass--black hole mass relation (with the horizontal grey dash-dotted line at 5 SMBH seed masses), the cosmic star formation rate density evolution $\dot{\rho}_\star (z)$, and the stacked $z = 0$ stellar mass profiles for haloes with $M_\mathrm{200c} = [10^{12}, 10^{13}]\,\msun$. Different colours represent four different models, as indicated in the top-right corner. In the left-hand and central panels, grey dashed lines denote the observational best-fit scaling relations of \citet{Marasco_et_al_2021} and \citet{Madau_Dickinson_2014}, respectively; these are shown for approximate guidance only, since we have made no attempt to reproduce observational selection functions. In the left-hand panel, small circles represent individual galaxies (for clarity only shown for the \texttt{Default} and \texttt{NoRepositioning} simulations), while solid lines and shaded bands trace the running medians and their $1\sigma$ uncertainties, respectively (omitted for \texttt{NoRepositioning} to highlight the extremely small galaxy-to-galaxy scatter). Without repositioning (yellow), there is no SMBH growth or AGN feedback at all. A more gradual drifting of SMBHs towards the galaxy centre (blue) allows substantial SMBH growth, but still affects star formation far less than with instantaneous repositioning (red), especially at $z \gtrsim 1$.}
	\label{fig:eagle_comparison}
\end{figure*}

In Fig.~\ref{fig:eagle_comparison}, we compare predictions from four variants of the \eagle{}-resolution simulations: \texttt{Default} (our baseline model; red), \texttt{NoRepositioning} (repositioning switched off; yellow), \texttt{DriftSpeed10kms} (instantaneous repositioning replaced by gradual drifting down the potential gradient at a fixed speed of 10 km/s; blue), and \texttt{NoAGN} (SMBH accretion, and hence also AGN feedback, switched off completely; green). From left to right, we show the relation between galaxy stellar mass $\mstar$ (the total mass of star particles bound to each galaxy within 30 proper kpc from its potential minimum) and black hole mass $M_\mathrm{BH}$ (defined as the subgrid $m_\mathrm{BH}$ of the most massive SMBH bound to each galaxy) at $z = 0$, the redshift evolution of the cosmic star formation rate density $\dot{\rho}_\star (z)$, and the stacked $z = 0$ stellar mass profiles for the 35 haloes with mass\footnote{$M_\mathrm{200c}$ is defined as the mass within the radius $r_\mathrm{200c}$ inside which the mean density equals 200 times the critical density of the Universe. Note that we select haloes based on their $M_\mathrm{200c}$ in a matched gravity-only simulation, to ensure that we compare the exact same objects between the different models.} $\log_{10}(M_\mathrm{200c}\,/\,\msun) = [12, 13]$, i.e.~towards the lower end of the range where AGN feedback is expected to have a significant impact \citep{Bower_et_al_2017,McAlpine_et_al_2017}. 

There are several key insights from this comparison. Most significantly, \emph{repositioning is an essential pre-requisite} for AGN feedback in our model. Without it, SMBHs experience essentially no growth at all, and both the global SFR history and stellar mass profiles are similar to the \texttt{NoAGN} model that explicitly disables AGN feedback. Perhaps counter-intuitively, the suppression of SMBH growth by disabling repositioning is even stronger than by switching off accretion completely (green). The reason for this is that in the latter case SMBHs can still grow by (repeated) mergers between seeds which, as a corollary, already accounts for essentially all the SMBH growth in galaxies with $\mstar \lesssim 3\times10^9\,\msun$. The gradual repositioning model \texttt{DriftSpeed10kms} (blue), on the other hand, allows significant SMBH growth, although with a noticeable shift in the $M_\mathrm{BH}$--$M_\star$ relation that we explore further in Section \ref{sec:comparison_drift}.

In terms of the cosmic SFR density $\dot{\rho}_\star (z)$, our simulations show a relatively constant offset of a factor $\approx\!\! 3$ between the \texttt{Default} and \texttt{NoRepositioning} models\footnote{We point out that the latter tracks the SFR density evolution of \texttt{NoAGN} almost perfectly, which rules out a significant impact of random run-to-run variations due to the non-deterministic nature of galaxy formation simulations at a low level (see also \citealt{Genel_et_al_2019} and \citealt{Keller_et_al_2019}).} at $z < 2$, with both runs converging at $z \gtrsim 4$ where the effect of AGN feedback is negligible. The gradual repositioning model \texttt{DriftSpeed10kms} (blue) represents a smooth transition between these two extreme cases: $\dot{\rho}_\star (z)$ closely tracks \texttt{NoAGN} until $z \approx 1.5$, and then gradually decreases more strongly, with an offset of only 0.1 dex from the instantaneous \texttt{Default} repositioning model at $z = 0$. We return to an interpretation of this feature below.

Finally, all three variations predict a similar stellar mass profile for massive $z = 0$ haloes, which are all markedly higher than in the \texttt{Default} model within the central few kpc, by a factor of 3--4. As an aside, we note that at least in terms of the stellar mass profiles, there is very little impact of AGN feedback on these galaxies beyond $\approx\!3$ kpc, comparable to their typical effective radii (e.g.~\citealt{Lange_et_al_2015,Trujillo_et_al_2020}).

As the careful reader may have noticed, our \texttt{Default} model predicts a cosmic SFR density that falls below the observed best-fit relation of \citet{Madau_Dickinson_2014} at $z \lesssim 3$. This mismatch is not too surprising, since we have made no detailed attempt to recalibrate our subgrid model following the updates with respect to \eagle{} (Section \ref{sec:model}); we will present such a recalibration in future work. For our purposes, the mismatch -- coincidentally almost exactly as large as the offset between the \texttt{Default} and \texttt{NoAGN} models -- is insignificant: the repositioning-related offsets are independent from it, at least to first order. To verify this explicitly, we have repeated the simulations shown in Fig.~\ref{fig:eagle_comparison} with a slightly adjusted SNe feedback parametrization that results in an excellent match of $\dot{\rho}_\star\, (z < 2)$ to the \citet{Madau_Dickinson_2014} relation (not shown). The offsets between the four model variants agree closely with those seen from our default parametrization as used in Fig.~\ref{fig:eagle_comparison} and in the rest of this paper.  

\begin{figure*}
	\includegraphics[width=2.1\columnwidth]{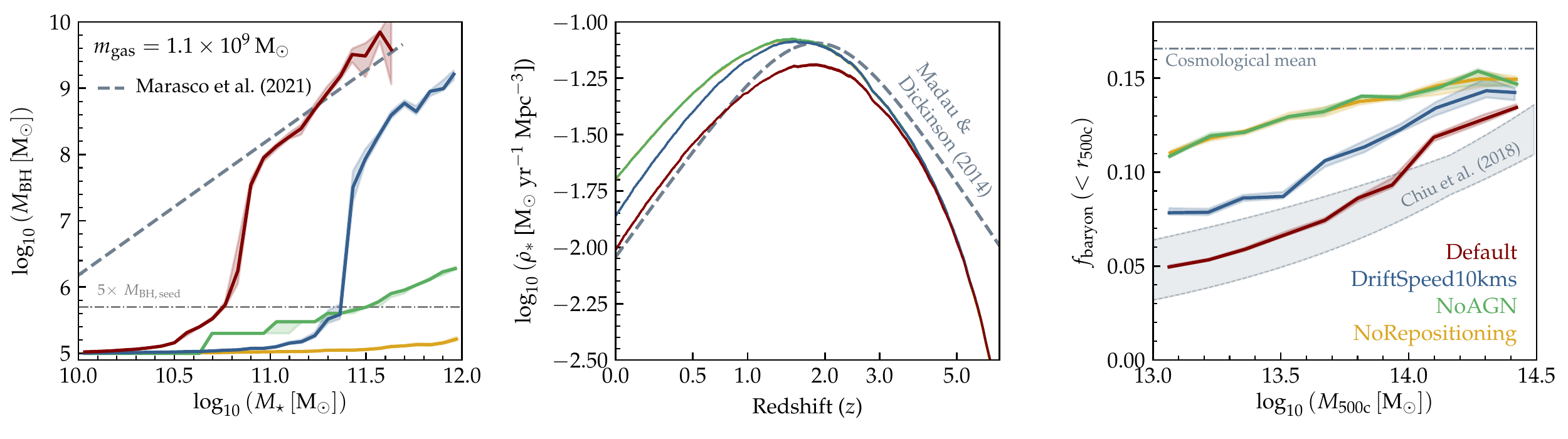}
	\caption{Comparison between different repositioning approaches for \bahamas{}-resolution simulations. The left-hand and central panels show predictions for the black hole mass--stellar mass relation and cosmic star formation rate density evolution, in analogy to Fig.~\ref{fig:eagle_comparison}. In the right-hand panel, the predicted baryon fractions of group/cluster haloes are plotted, with the observational relation of \citet{Chiu_et_al_2018} indicated as a grey band for approximate guidance.  Despite differences in detail, the conclusions are qualitatively consistent with those from our \eagle{} resolution simulations: SMBH repositioning has a dramatic effect on black hole growth, star formation, and the baryon content of massive haloes.}
	\label{fig:bahamas_comparison}
\end{figure*}

To explore the sensitivity of these results to resolution and the limited volume of the \eagle-resolution simulations, we show the analogous comparison for the \textsc{bahamas}-resolution runs in Fig.~\ref{fig:bahamas_comparison}; since kpc-level galaxy structures are not resolved here, we omit the stellar profiles and instead compare the baryon fractions of group/cluster-scale haloes that are missing in the small volume of the \eagle{}-resolution simulations. Qualitatively, offsets between the different repositioning models are the same as for the \eagle{}-resolution runs despite their three orders of magnitude difference in particle mass; the impact on the group/cluster baryon fractions mimics that on $\dot{\rho}_\star(z)$. In detail, the gradual drift model (\texttt{DriftSpeed10kms}, blue) deviates more strongly from the \texttt{Default} (instantaneous repositioning) approach than at \eagle{} resolution, with SMBH growth strongly suppressed up to galaxy masses of $\mstar \approx 2\times 10^{11}\,\msun$; the impact on the cosmic SFR history is also somewhat larger (0.15 dex difference at $z = 0$). As above, \texttt{NoRepositioning} leads to negligible SMBH growth at any stellar mass and a star formation history that is nearly identical to \texttt{NoAGN}.

\subsection{Gradual instead of instantaneous repositioning}
\label{sec:comparison_drift}
\begin{figure*}
	\includegraphics[width=2.1\columnwidth]{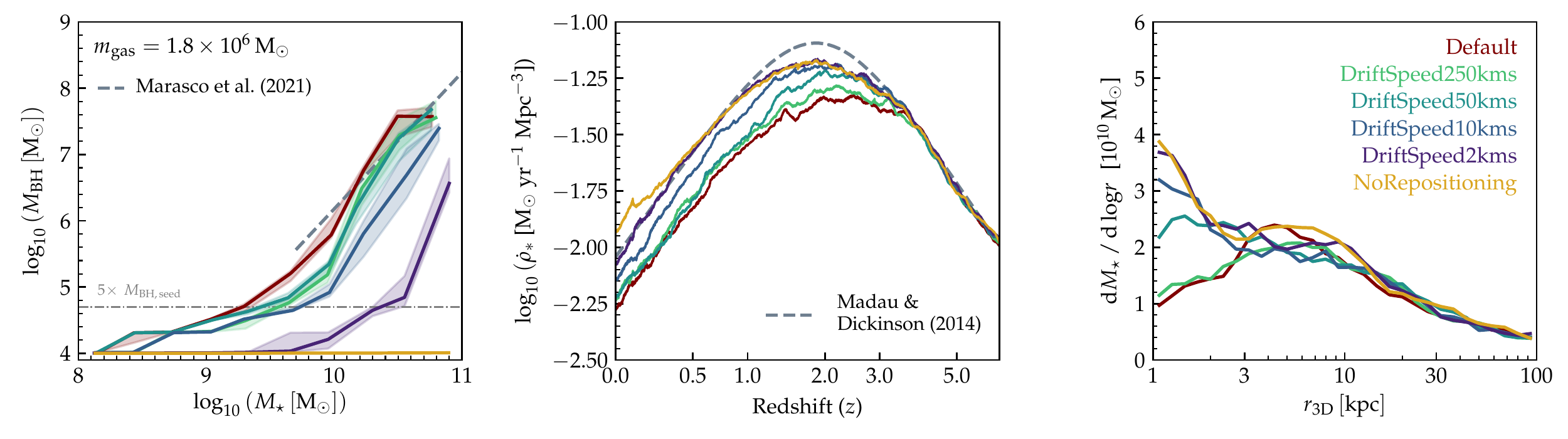}
	\caption{Similar to Fig.~\ref{fig:eagle_comparison}, but comparing \eagle{}-resolution runs with four different drift speeds (shades of blue/green) to the \texttt{Default} (red) and \texttt{NoRepositioning} (yellow) runs. Lower SMBH drift speeds lead to systematically weaker black hole growth, later onset of AGN feedback, and higher central stellar densities in massive galaxies. Qualitatively similar trends are found in the \bahamas{} resolution simulations (not shown).}
	\label{fig:eagle_vrepos}
\end{figure*}

Although we have argued above that a drift speed of $\sim$10 km s$^{-1}$ is approximately consistent with typical SMBH sinking time scales under (gravitational) dynamical friction, the precise value is highly uncertain. In Fig.~\ref{fig:eagle_vrepos}, we show the same predictions as in Fig.~\ref{fig:eagle_comparison} but now comparing four (\eagle{}-resolution) simulations with different drift speeds $v_\mathrm{drift}$ (increasing by successive factors of 5 from 2 to 250 km s$^{-1}$). There are clear trends with $v_\mathrm{drift}$ in all three panels, with \texttt{DriftSpeed2kms} closest to \texttt{NoRepositioning} and \texttt{DriftSpeed250kms} closest to \texttt{Default}. From the central panel, AGN in \texttt{DriftSpeed2kms} only affect the simulation at $z \lesssim 0.5$, whereas for \texttt{DriftSpeed250kms}, the effect on star formation is essentially the same as in the instantaneous-repositioning \texttt{Default} run at $z \lesssim 1.5$. This gradually earlier onset (and earlier full effect) of AGN feedback with higher $v_\mathrm{drift}$ can be understood in terms of the shorter time it takes for SMBHs to reach the galaxy centre. It corresponds to a gradually lower central stellar mass in massive haloes, while star formation beyond 3 kpc is not significantly affected.  Although we only show the \eagle{}-resolution simulations here, we have verified that the \bahamas{}-resolution runs show the same qualitative behaviour, consistent with our expectations from Fig.~\ref{fig:bahamas_comparison}.

\begin{figure}
	\includegraphics[width=\columnwidth]{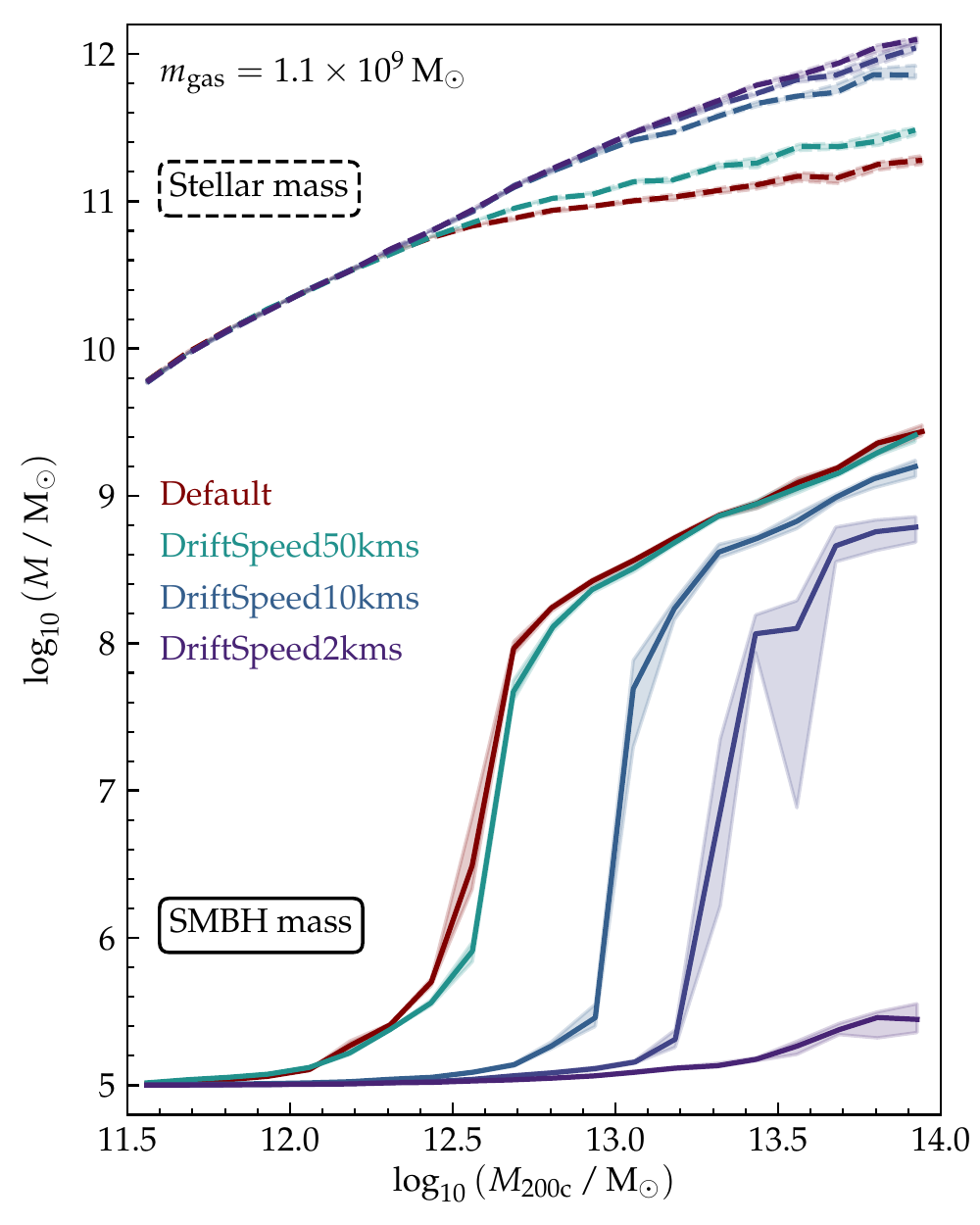}
	\caption{The effect of limiting the reposition speed on the median stellar mass (dashed lines, top) and SMBH mass (solid lines, bottom), as a function of halo mass $M_\mathrm{200c}$. Lines in shades of blue and green represent four models with a fixed repositioning drift speed at \bahamas{} resolution, which are compared to the \texttt{Default} model (instantaneous repositioning, red). Shaded bands indicate $1\sigma$ bootstrapping uncertainties on the median. Reducing the repositioning speed leads to systematically higher stellar masses, shifts the rapid growth scale of SMBHs to higher halo masses, and reduces the masses of self-regulating SMBHs in massive haloes.}
	\label{fig:bahamas_vdrift}
\end{figure}

In the figures above, the $M_\mathrm{BH}$--$M_\star$ relation of models with gradual repositioning is systematically offset from the instantaneous-repositioning \texttt{Default} model. To understand whether this is due to an impact on the growth of stellar mass, SMBHs, or both, we plot in Fig.~\ref{fig:bahamas_vdrift} the separate relations between both quantities and halo mass $M_\mathrm{200c}$. We show results from the \bahamas{}-resolution simulations with the \texttt{Default} model and four variations with constant drift speed $v_\mathrm{drift}$ between 2 and 50 km s$^{-1}$ (\texttt{DriftSpeed250kms} is omitted for clarity). Qualitatively consistent behaviour is seen for the equivalent \eagle{}-resolution runs (not shown).

Making the SMBH repositioning non-instantaneous clearly affects both stellar and SMBH masses. Consistent with the increase in the cosmic star formation rate density seen in Fig.~\ref{fig:eagle_vrepos}, slower repositioning (lower $v_\mathrm{drift}$) leads to systematically higher stellar masses. This difference increases steadily towards more massive haloes\footnote{There is no difference in stellar mass between the different models at $\log_{10}\,(M_\mathrm{200c}\,/\,\msun) < 12.4$, even though the SMBH masses already show a clear offset at this point. This is a consequence of the low resolution of the simulations shown here: due to the large amount of energy that is required to heat even a single particle, AGN feedback can only occur once the SMBH has accreted $\sim$10$^7\,\msun$ of gas.}, and reaches up to an order of magnitude for $M_\mathrm{200c} \approx 10^{14}\,\msun$. While this contributes to the offset in the $M_\mathrm{BH}$--$M_\star$ relation (a shift to the right for slower repositioning), there are also two clear direct effects on SMBH growth itself. Firstly, the halo mass scale at which SMBHs experience rapid growth \citep{Bower_et_al_2017} shifts up as repositioning is slowed down, by almost a factor of 10 between $v_\mathrm{drift} =$ 50 and 5 km s$^{-1}$. We defer a detailed exploration of this effect to future work, but note that one plausible mechanism is that at larger halo-centric distances (where SMBHs are typically found if repositioning is slow, see Section \ref{sec:analysis_positions}), gas densities are lower, so that SN feedback remains effective, and SMBH growth therefore suppressed \citep{Bower_et_al_2017} up to larger halo masses. In addition, the lower gas densities may also directly reduce SMBH growth. Alternatively, SMBHs may remain closer to the centre of more massive haloes even with slower repositioning, or differences in the build-up of haloes over cosmic time may play a role (as seen above, slower repositioning leads to a later onset of effective AGN feedback).

Secondly, the relation for massive haloes -- i.e.~those in which SMBHs have grown substantially, to the right of the aforementioned rapid growth point -- is systematically lower for lower $v_\mathrm{drift}$. As discussed by \citet[see also \citealt{Bower_et_al_2017}]{Booth_Schaye_2009}, SMBHs in this regime are in a state of self-regulated growth, with lower masses originating from more efficient feedback, rather than the available gas. At first, this may seem to contradict what we found above, namely that gradual drifting of SMBHs down the potential gradient (especially at low $v_\mathrm{drift}$) leads to a reduced impact of AGN feedback on star formation and baryon content compared to instantaneous repositioning. A plausible explanation is that smaller cooling losses in the less dense gas outside the galaxy centre \citep{DallaVecchia_Schaye_2012} increase the local efficiency of AGN feedback when SMBHs are tied less strongly to the centre of their host galaxy, but that its galaxy-wide impact is diminished because the energy is injected away from the dense, star-forming core. In detail, the situation is likely even more complex, as hinted at by e.g.~the qualitatively different offset between the \texttt{Default} and \texttt{DriftSpeed50kms} models for stellar and SMBH mass.

\subsection{The effect of a velocity threshold for repositioning}
\begin{figure}
	\includegraphics[width=\columnwidth]{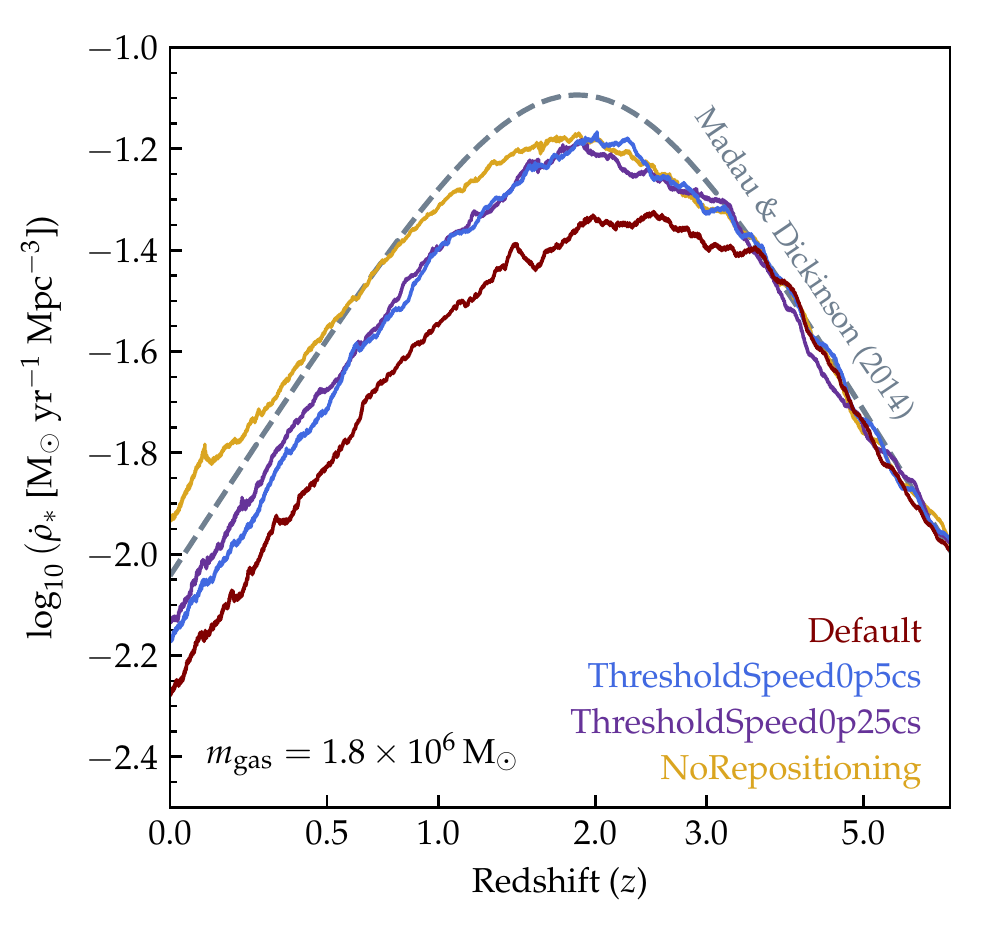}
	\includegraphics[width=\columnwidth]{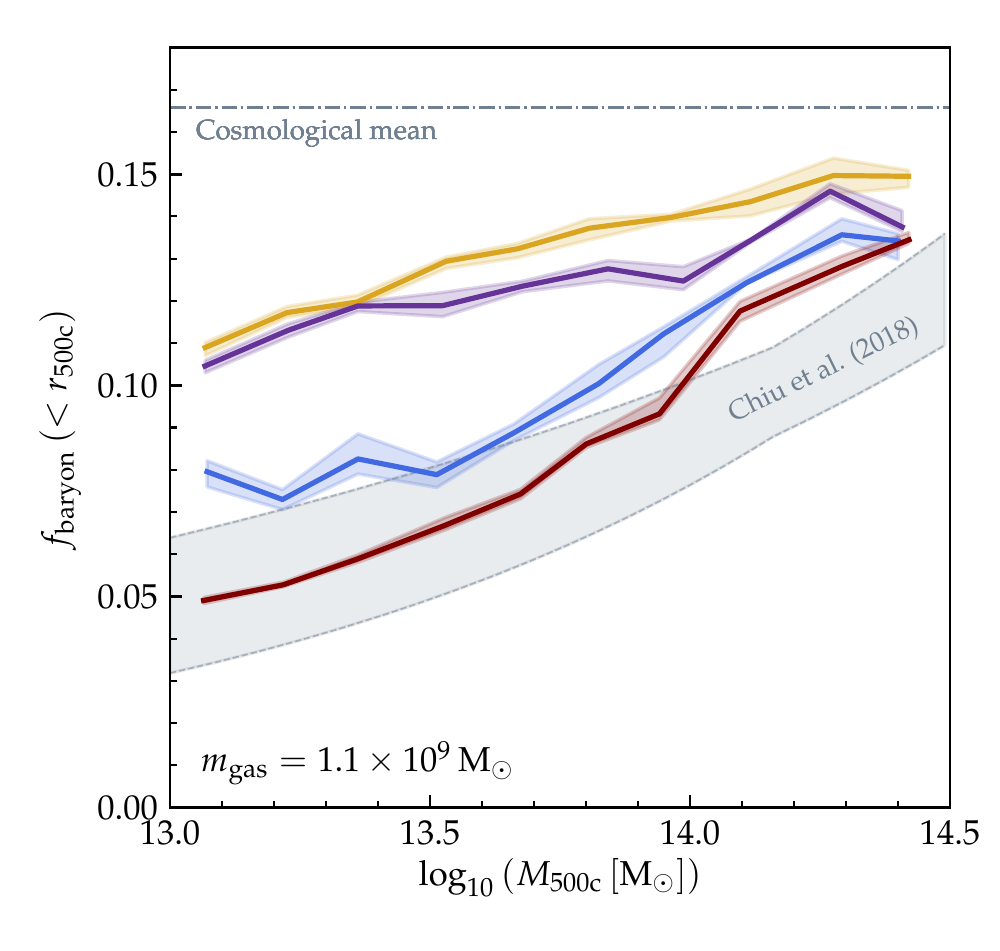}
	\caption{SFR history (top panel, \eagle{} resolution) and group/cluster baryon fractions (bottom panel, \bahamas{} resolution) for two simulations that limit repositioning to particles moving slowly with respect to the SMBH (blue, purple), compared to repositioning onto any particle (red) and no repositioning (yellow). The velocity threshold reduces the efficiency of AGN feedback considerably, at both resolution levels.}
	\label{fig:velcut}
\end{figure}

Finally, we test the impact of limiting SMBH repositioning to slowly moving neighbour particles, as was done in some previous simulations (see above). In Fig.~\ref{fig:velcut}, we compare the cosmic SFR histories and group/cluster baryon fractions (for \eagle{} and \bahamas{} resolution simulations, respectively) for the two models that include this velocity threshold, \texttt{VelocityThreshold0p25cs} and \texttt{VelocityThreshold0p5cs}.  

It is evident that the velocity threshold, intended only as a guard against accidental jumps of SMBHs between galaxies, has a significant effect on the baryon distribution and star formation in the simulation. Throughout the redshift range where AGN are important ($z \lesssim 4$), the velocity threshold models predict far higher star formation rates, by more than 0.2 dex (also at the lower \bahamas{} resolution, which is not shown). The onset of AGN feedback is also delayed, to $z \approx 2$ in our simulations (the exact value is likely sensitive to other subgrid parameters, as well as the simulation volume). At the lower resolution of \bahamas{}, the more extreme threshold (0.25 $c_\mathrm{sound}$, purple) almost completely prevents AGN feedback and leads to baryon fractions close to \texttt{NoRepositioning} at $z = 0$, whereas the \eagle{} resolution runs show only a very small difference between thresholds of 0.25 and 0.5 $ c_\mathrm{sound}$. We have verified that a fixed velocity threshold of 30 km s$^{-1}$, independent of $c_\mathrm{sound}$, leads to similarly inefficient repositioning.

Like the other repositioning-related offsets, the impact of the velocity threshold may be mitigated by adjusting other subgrid parameters. We also reiterate that there is a subtle difference between our simulations and the actual \eagle{} and \bahamas{} runs in that those also allowed repositioning onto star and dark matter particles. However, it is unclear whether such parameter changes could fully compensate the velocity threshold in SMBH repositioning. 

The reason for the large effect of the repositioning velocity threshold is that gas near the centre of galaxies is typically rotation-supported (at least on the scales resolved by our simulations), so that typically $v \gg c_\mathrm{sound}$; this is particularly true in galaxies with high central gas fractions and star formation rates. SMBHs are therefore unable to reposition until either the ambient gas temperature (and hence $c_\mathrm{sound}$) increases significantly -- but as we have seen, AGN feedback that could cause such a temperature increase itself requires repositioning -- or a temporary disruption to the velocity structure of the gas through e.g.~a (galaxy) merger. Our results suggest that this (unintended) side-effect of a repositioning velocity threshold should preclude its use in future simulations and warrant careful interpretation of results from simulations that did include it.

\section{How does black hole repositioning affect galaxies?}
\label{sec:analysis}

We have shown above that SMBH repositioning (or an equivalent way of ensuring that they can sink towards the local potential minimum) is an essential prerequisite for efficient AGN feedback in intermediate- and low-resolution simulations; furthermore, repositioning is not just an ``on-off'' switch as the details of its implementation matter significantly. We now investigate \emph{why} it is so important: we first test its impact on the positions and merging of SMBHs (Sec.~\ref{sec:analysis_positions}), and then on their gas accretion rates (Sec.~\ref{sec:analysis_gas}). 

\begin{figure}
	\includegraphics[width=\columnwidth]{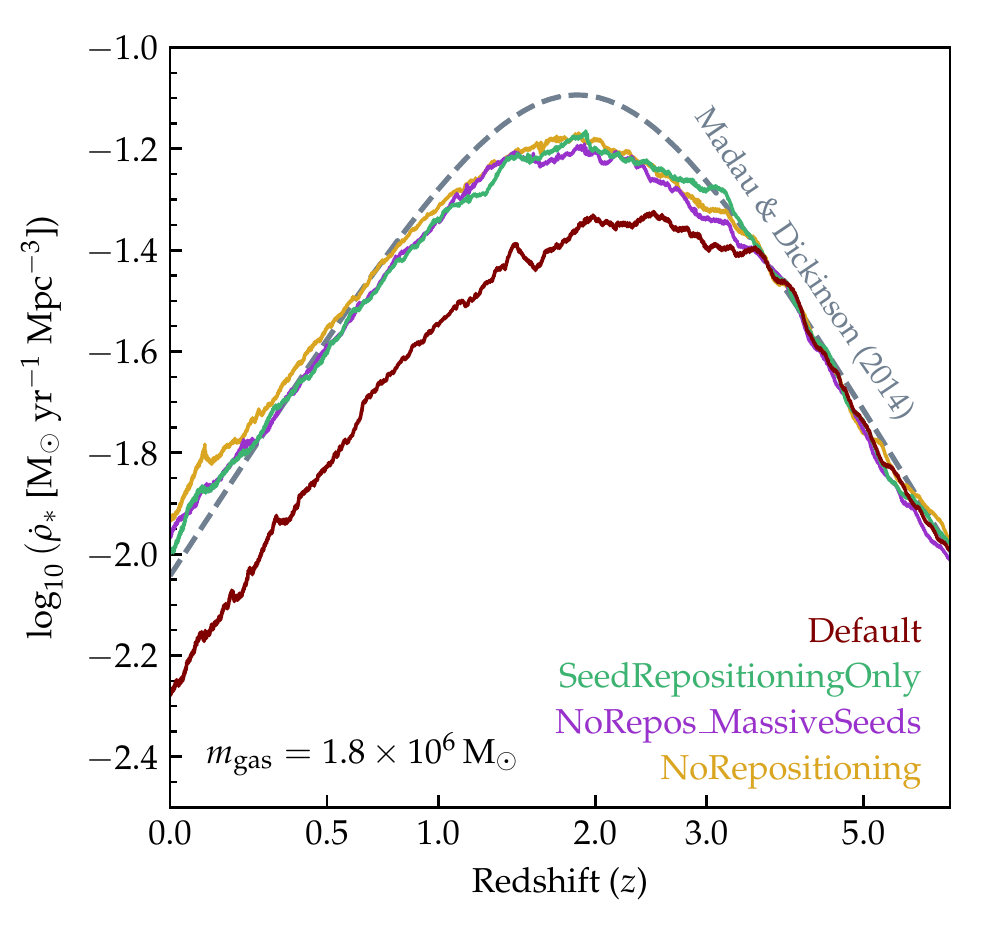}
	\includegraphics[width=\columnwidth]{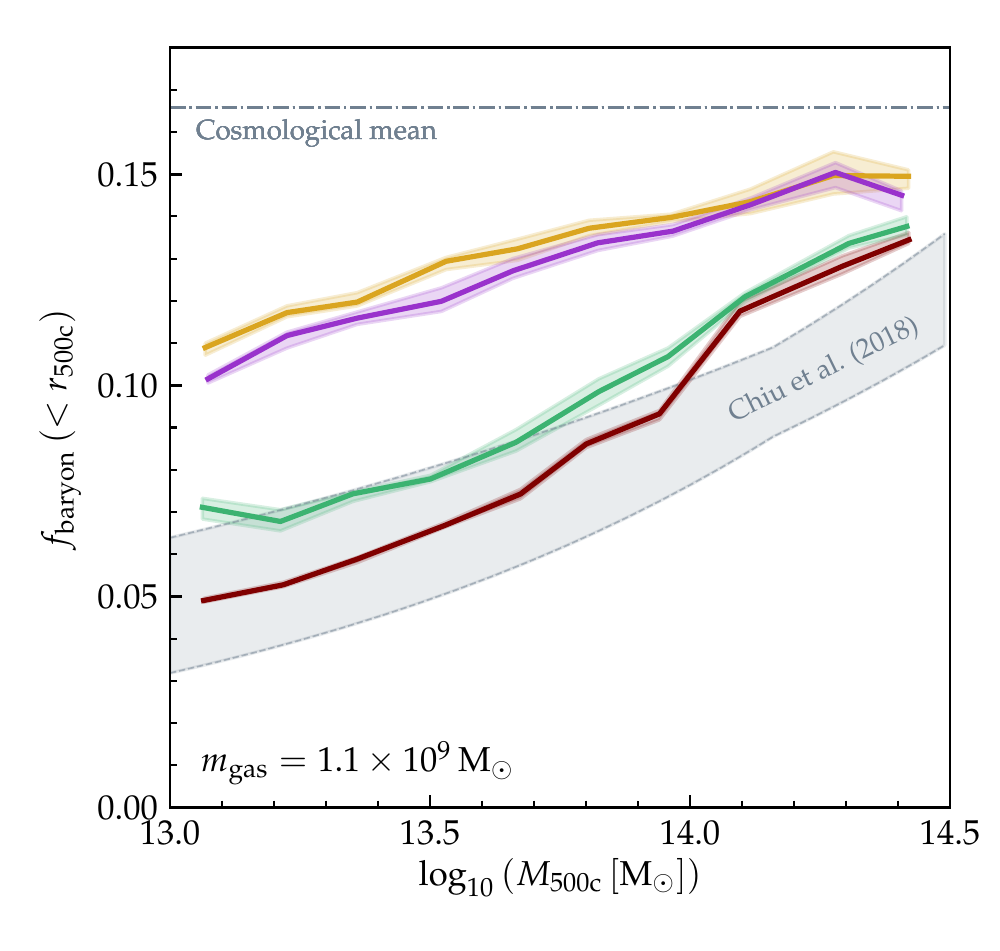}
	\caption{SFR history (top panel, \eagle{} resolution) and group/cluster baryon fractions (bottom panel, \bahamas{} resolution) for models \texttt{SeedRepositioningOnly} (where repositioning is only enabled while SMBHs have grown by less than 20 per cent, green) and \texttt{NoRepos\_MassiveSeeds} (no repositioning but SMBHs are seeded at $100\times$ higher mass, magenta), compared to repositioning onto any particle (red) and no repositioning (yellow). While neither of these two intermediate approaches show much of an effect in the top panel, repositioning SMBHs immediately after seeding accounts for most of its impact on the baryon content of massive haloes at \bahamas{} resolution.}
	\label{fig:growth_repos}
\end{figure}

As a preliminary step, we test \emph{at which stage} of SMBH growth repositioning is important: to allow an initial migration to the galaxy centre after seeding, to help SMBHs acquire sufficient mass for Bondi-Hoyle-Lyttleton accretion to become efficient, or throughout their life times? For this, we compare the models \texttt{SeedRepositioningOnly}, in which SMBHs are only repositioned for a limited time after seeding (until their mass has grown by at least 20 per cent, i.e.~typically until the first SMBH merger) and \texttt{NoRepos\_MassiveSeeds}, in which repositioning is switched off but SMBHs are seeded at $100\times$ higher mass (i.e.~$10^6\,\msun$ and $10^7\,\msun$ for the \eagle{} and \bahamas{} resolution simulations, respectively), in Fig.~\ref{fig:growth_repos}.

At \eagle{} resolution (top panel), neither of these two deviates significantly from \texttt{NoRepositioning}; AGN feedback remains insignificant throughout the simulation. The situation is less clear for the baryon fractions of massive haloes in the \bahamas{} resolution runs; while \texttt{NoRepos\_MassiveSeeds} is still close to \texttt{NoRepositioning}, the \texttt{SeedRepositioningOnly} model predicts baryon fractions that are only $\lesssim$\,50 per cent higher than in the \texttt{Default} (continuous) repositioning approach. From inspecting individual growth tracks of SMBHs in these massive haloes (not shown), we have found that they grow much of their mass (and inject the bulk of their feedback energy) within a short ($\lesssim 1$ Gyr) period of time, starting from near seed mass. This is likely too rapid for the SMBHs to wander significantly away from the galaxy centre, even if repositioning ends as soon as this growth phase begins. The cosmic SFR density evolution in \texttt{SeedRepositioningOnly} at \bahamas{} resolution on the other hand, which is dominated by less massive haloes, is more strongly offset from \texttt{Default} (not shown). We therefore conclude that SMBH repositioning plays a particular role directly after seeding, but that it clearly influences black hole growth and AGN feedback also at later stages.

\subsection{BH positions and mergers}
\label{sec:analysis_positions}
\begin{figure*}
	\includegraphics[width=2.1\columnwidth]{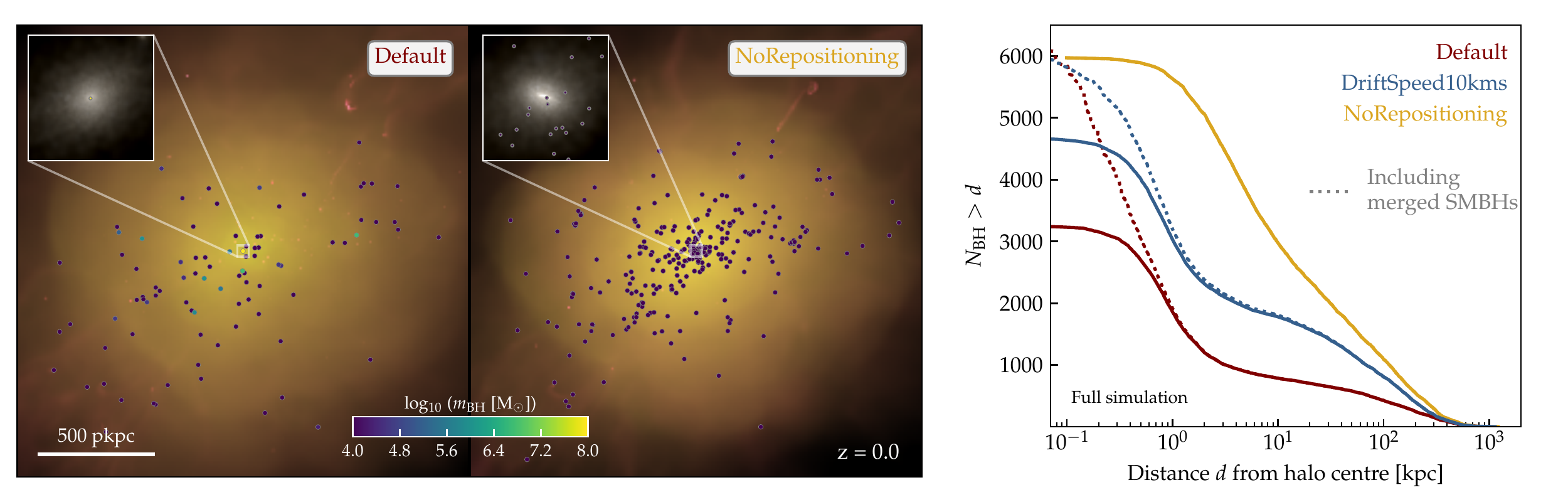}
	\caption{Projected SMBH positions (circles, colour-coded by $m_\mathrm{BH}$) are plotted over the corresponding gas density-temperature map (brightness representing surface density and hue temperature, increasing from pink to yellow) of the most massive $z = 0$ halo ($M_\mathrm{200c} = 1.7\times 10^{13}\,\msun$) in the 25 Mpc \eagle{}-resolution simulations with the \texttt{Default} (left) and \texttt{NoRepositioning} (centre) models, respectively. The insets show synthetic \textit{gri} images of the central galaxy with a side length of 50 kpc. Right: cumulative radial distribution function of SMBHs with respect to the potential minimum of their host haloes in the full simulation volumes. SMBHs are spread throughout the halo irrespective of repositioning, but there is a far larger number of them in the \texttt{NoRepositioning} model, especially near the halo centre.}
	\label{fig:bh_positions}
\end{figure*}

The spatial distribution of SMBHs within their host halo is what is most directly affected by repositioning. On the left-hand side of Fig.~\ref{fig:bh_positions}, we show a visualization of its effect on the most massive halo in the \eagle{} resolution simulations ($M_\mathrm{200c} = 1.7 \times 10^{13}\,\msun$ in the DM-only version); we compare this halo in the \texttt{Default} (left) and \texttt{NoRepositioning} (right) runs at $z = 0$.

A number of features in this comparison are worth pointing out. Perhaps surprisingly, the large-scale extent of the halo of SMBHs is \emph{not} significantly different between the two models; SMBHs are found out to radii of $>\!\!500$ kpc in both cases. Closer to the centre, however, there is a marked difference (see the zoom-in insets): repositioning causes the difference between more than a dozen SMBHs distributed throughout the central 50 kpc and a single one sitting at the very centre. Similarly, repositioning also reduces the number of SMBHs at larger radii. Consistent with Fig.~\ref{fig:eagle_comparison}, none of the SMBHs in the \texttt{NoRepositioning} run have grown significantly beyond the seed mass of $m_\mathrm{seed} = 10^4\,\msun$ (dark purple colour), whereas the \texttt{Default} model has a number of SMBHs with $m_\mathrm{BH} \gg m_\mathrm{seed}$ (green/yellow points). Finally, we point out that the gas and stellar structure of the galaxy halo as shown in the background images is significantly different, as expected from our results in the previous section: without repositioning, both components show a strong central concentration, whereas the \texttt{Default} model produces a noticeably more diffuse gas halo and central group galaxy, at a level that significantly exceeds the expected variation from run-to-run stochasticity.

These differences are summarized in a more quantitative fashion by the plot on the right-hand side of Fig.~\ref{fig:bh_positions}, which shows the cumulative radial distribution function of SMBHs within all host haloes of the respective simulation at $z = 0$. In addition to the two runs depicted on the left-hand side, we also include the \texttt{DriftSpeed10kms} model here; recall that this drifts SMBHs more gradually towards the local potential minimum than the instantaneous repositioning of our \texttt{Default} model. While all three converge at the far end (with maximum SMBH distances of $\approx\!\!500$ kpc), weaker repositioning leads to a larger number of SMBHs at all smaller radii; in total, the \texttt{NoRepositioning} and \texttt{DriftSpeed10kms} runs contain a factor of $\approx$2 and $\approx$1.5 times more SMBHs than \texttt{Default} at $z = 0$.

The seeding of SMBHs is largely independent of repositioning, so that these differences suggest a (strong) effect on SMBH merger rates. This is not surprising, given that our SMBH merger criterion requires BHs to approach one another closely in phase-space, precisely what repositioning achieves, and also implies that simulation predictions for gravitational waves from SMBH mergers (e.g.~\citealt{Salcido_et_al_2016}) are strongly affected by repositioning. The dotted lines in the right-hand panel of Fig.~\ref{fig:bh_positions} account for such mergers by weighting each SMBH by the number of seeds it contains, i.e.~the total number of mergers that have contributed to its growth (for \texttt{NoRepositioning}, the dotted and solid blue lines lie on top of each other). As expected, these lines do all converge at small radii, confirming that the total number of SMBHs that were seeded is similar across the models. The comparison between these merger-corrected distributions then clearly reveals the expected impact of repositioning; with increasing efficiency, SMBHs are located closer to the halo centre (with the medians ranging between 10 kpc for \texttt{NoRepositioning} and 0.5 kpc for \texttt{Default}). For the halo shown on the left-hand side of Fig.~\ref{fig:bh_positions}, there are a total of 144 SMBH--SMBH mergers in the \texttt{Default} run, but not a single one in the \texttt{NoRepositioning} variant. Although only the distributions at $z = 0$ are shown for clarity, we have verified that a qualitatively consistent picture is seen at $z = 1$ and $z = 2$.

\begin{figure}
	\includegraphics[width=\columnwidth]{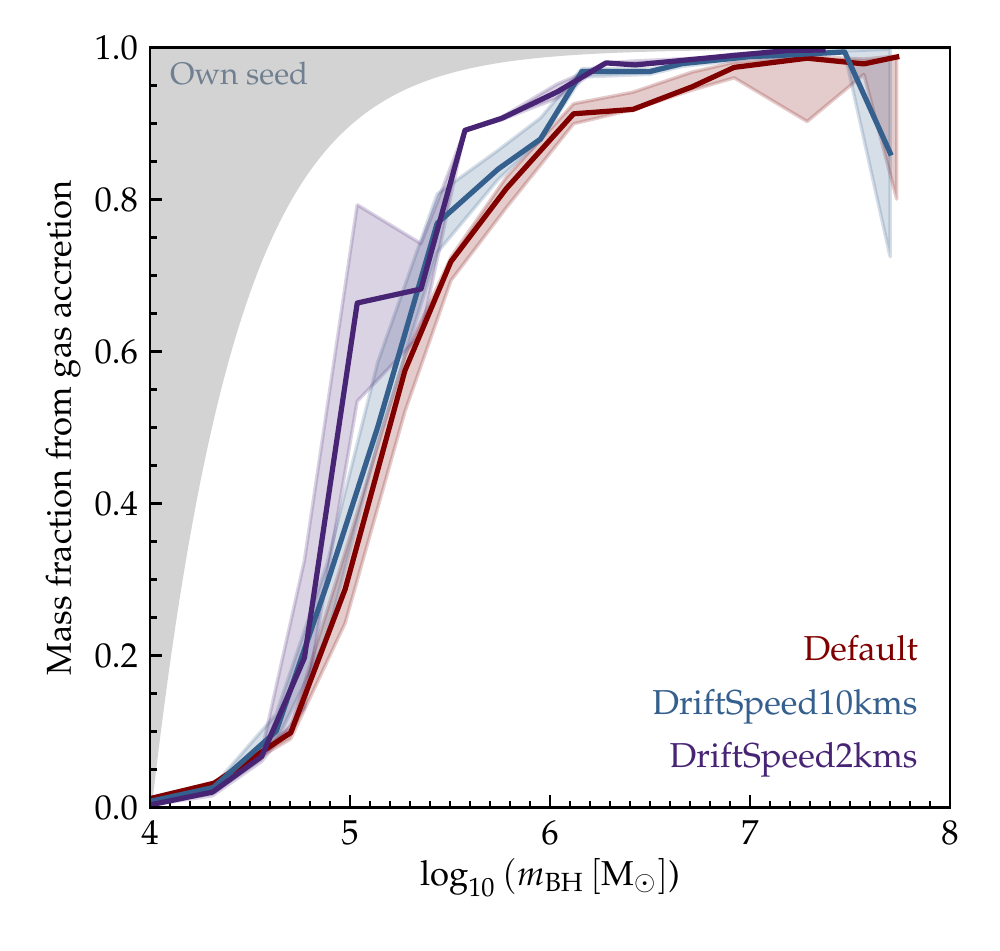}
	\caption{The fraction of SMBH mass $m_\mathrm{BH}$ gained through direct gas accretion, rather than from mergers with other SMBHs, for models \texttt{Default}, \texttt{DriftSpeed2kms}, and \texttt{DriftSpeed10kms}. Solid lines represent running medians, shaded bands $1\sigma$ uncertainties from bootstrap resampling. The grey area in the top-left corner indicates the region that is inaccessible due to the seed mass of the main progenitor itself. Although mergers are suppressed without efficient repositioning (Fig.~\ref{fig:bh_positions}), the effect on the growth of massive SMBHs is modest, because these gain most of their mass via gas accretion.}
	\label{fig:bh_facc}
\end{figure}

Since repositioning increases the number of SMBH mergers, a natural next question is to what extent this alone can explain the different SMBH masses and -- since Bondi-Hoyle-Lyttleton accretion depends strongly on $m_\mathrm{BH}$ -- AGN feedback efficiencies between the different models\footnote{For clarity, we emphasize that we here only refer to (subgrid) mergers between different SMBHs. Galaxy mergers are also expected to have a significant impact on the gas-accretion driven growth of SMBHs, as discussed extensively in the literature (e.g.~\citealt{Hopkins_et_al_2008,McAlpine_et_al_2018,Lapiner_et_al_2021}).}. While it is evident from Fig.~\ref{fig:eagle_comparison} that mergers play \emph{some} role in the growth of SMBHs -- they are the only option for SMBH growth in the \texttt{NoAGN} run -- we show in Fig.~\ref{fig:bh_facc} that they are nevertheless strongly sub-dominant to (direct) gas accretion for $m_\mathrm{BH} \gtrsim 10^6\,\msun$; even in the \texttt{Default} model, around 95 per cent of the mass in SMBHs with $m_\mathrm{BH} = 10^7\,\msun$ comes from gas accretion onto the main progenitor, rather than from merged SMBHs. Although less efficient repositioning (and hence fewer SMBH mergers) increases the fraction of accretion-driven growth yet further, there is therefore little room for substantial differences between the repositioning models in this respect\footnote{We point out, however, that Fig.~\ref{fig:bh_facc} only considers the total mass growth over the entire simulation; it is well conceivable that SMBH mergers play a larger role in the early growth of massive SMBHs. We defer a detailed investigation of this question to future work.}.

\subsection{Ambient gas properties}
\label{sec:analysis_gas}

We have seen above that weak or no repositioning leads to drastically reduced SMBH growth and AGN feedback, even when black holes are seeded at 100 times higher mass to compensate for any suppressed early growth (magenta line in Fig.~\ref{fig:growth_repos}). We have also shown that SMBH mergers contribute negligibly to the growth of those massive SMBHs that are chiefly responsible for AGN feedback. All of this suggests that gas accretion itself is strongly dependent on repositioning.

\begin{figure}
	\includegraphics[width=\columnwidth]{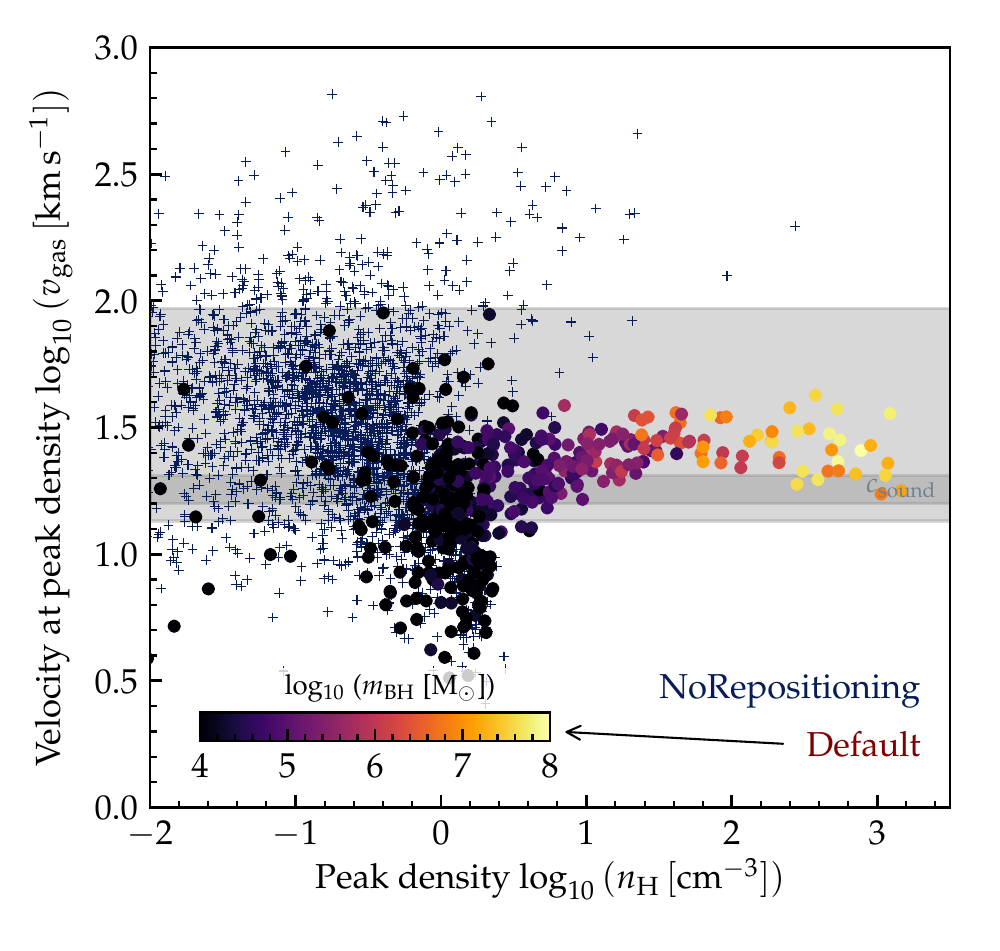}
	\caption{Peak gas density around SMBHs in haloes with $M_\mathrm{200c} > 10^{11}\,\msun$ at $z = 0$ ($x$-axis) plotted vs.~the ambient gas speed at this point ($y$-axis). Both quantities are calculated as medians over a 500 Myr interval; see text for details. The \texttt{NoRepositioning} model is shown as dark blue crosses, the \texttt{Default} model as circles whose colour indicates the final SMBH mass $m_\mathrm{BH}$. Dark and light grey horizontal bands cover the central 20 and 68 per cent of the ambient sound speed in the latter model. Without repositioning, fewer SMBHs encounter the high densities ($n_\mathrm{H} \gtrsim 10\,\mathrm{cm}^{-3}$) required for significant growth; where they are encountered, the velocity relative to the gas is too high to permit significant accretion.}
	\label{fig:bh_densvel}
\end{figure}

A direct comparison between accretion rates in different repositioning models is, however, non-trivial: its strong dependence on $m_\mathrm{BH}$ makes comparisons across SMBH masses meaningless, but as shown in Figs.~\ref{fig:eagle_comparison} and \ref{fig:bahamas_comparison}, SMBHs of the same mass live in very different large-scale environments in different models, if they exist at all. Instead, we compare in Fig.~\ref{fig:bh_densvel} the two \emph{external} variables of the Bondi-Hoyle-Lyttleton accretion formula, i.e.~ambient gas density and velocity. These could be compared at fixed redshifts across models, but since SMBH accretion is expected to be dominated by rapid growth phases \citep{McAlpine_et_al_2018}, such a comparison would likely be dominated by `uninteresting' SMBHs that are relatively quiescent at the time of consideration.

Instead, we make use of the high-time-resolution outputs that we store for SMBH properties (see Section \ref{sec:sim_ics}) and determine for each individual SMBH the peak sustained gas density that it ever experienced. More specifically, we calculate the median density over a sliding window of 500 Myr between its seeding and $z = 0$ and record the maximum of these. This quantity is plotted on the $x$-axis of Fig.~\ref{fig:bh_densvel}, against the median ambient gas velocity over the same time interval. For clarity, only SMBHs in haloes with $M_\mathrm{200c} > 10^{11}\,\msun$ at $z = 0$ are shown; we compare the most strongly differing models \texttt{Default} and \texttt{NoRepositioning} from the \eagle{}-resolution simulations.

Two clear differences arising from SMBH repositioning are apparent. Firstly, peak densities in the \texttt{Default} models stretch to far higher values than in \texttt{NoRepositioning}, with $n_\mathrm{H} > 1000\,\mathrm{cm}^{-3}$ in the most extreme cases. There is a clear and strong correlation between peak density and ($z = 0$) $m_\mathrm{BH}$: the most massive ones are those that have experienced the highest densities. In \texttt{NoRepositioning}, on the other hand, only around a dozen SMBHs have peak densities exceeding $10\,\mathrm{cm}^{-3}$, and only two have $n_\mathrm{H} \gtrsim 100\,\mathrm{cm}^{-3}$. This is consistent with the larger halo-centric radii of SMBHs in this model (Fig.~\ref{fig:bh_positions}), since gas density decreases with radius.

Secondly, those few SMBHs that do encounter high densities without repositioning do so at much higher velocity $v_\mathrm{peak}$ than in the \texttt{Default} model. In the latter, $v_\mathrm{peak}$ is typically below 30 km s$^{-1}$, only marginally above the characteristic sound speed of the ambient gas. Without repositioning, on the other hand, moderately high peak densities tend to correspond to very high $v_\mathrm{peak} \gtrsim 100\,\mathrm{km}\, \mathrm{s}^{-1}$. This is well above the sound speed and therefore controls the denominator in the expression for the Bondi-Hoyle-Lyttleton accretion rate (equation \ref{eq:bondi}). Since $\dot{m}_\mathrm{BH} \propto v_\mathrm{gas}^{-3}$, this factor $\sim$10 difference in velocity corresponds to a $\sim$1000$\times$ suppression of $\dot{m}_\mathrm{BH}$, or $\sim$10$^5\times$ when combined with the difference in density ($\dot{m}_\mathrm{BH} \propto n_\mathrm{H}$).

Finally, we note that both models contain a large number of SMBHs that never experience (sustained) densities above a few cm$^{-3}$. These also show a significant offset between the two models, with a tendency for lower peak density and higher peak velocity without repositioning. In combination, this prevents essentially any significant gas accretion without repositioning.

\section{Conclusions}
\label{sec:summary}
Large-scale cosmological hydrodynamic simulations currently require simplistic ``repositioning'' prescriptions to mimic the effect of unresolved dynamical friction on supermassive black holes (SMBHs) and their AGN feedback. Unlike the prescriptions for black hole seeding, gas accretion, and AGN feedback, the effects of repositioning have not yet been investigated systematically. We have tested the impact of this repositioning on the growth of SMBHs and its associated effect on star formation and the distribution of baryons within massive haloes in a series of simulations ranging from the $\sim$10$^6\,\msun$ resolution characteristic of current galaxy formation simulations to the $\sim$10$^9\,\msun$ resolution used to model large-scale structure on cosmological scales. Gas accretion onto SMBHs is modelled with the Bondi-Hoyle-Lyttleton formula, starting from seed masses far below the SMBH mass scales of interest. Our main conclusions may be summarized as follows:

\begin{enumerate}
	\item Repositioning (or an equivalent explicit modelling of dynamical friction on SMBHs) is a necessary prerequisite for AGN feedback. SMBHs hardly grow in its absence, not even through mergers, with strong impacts on star formation, central stellar masses, and baryon fractions of massive haloes. Despite quantitative differences, this statement is qualitatively insensitive to resolution (Figs.~\ref{fig:eagle_comparison} and \ref{fig:bahamas_comparison}).

	\item Replacing the instantaneous repositioning of SMBHs down the gravitational potential gradient, as is done in most contemporary simulations (e.g.~\eagle{}, \textsc{Auriga}, \bahamas{}, \textsc{IllustrisTNG}, \textsc{Fable}, \textsc{Simba}, \textsc{Artemis}), with a more gradual drift of $\lesssim 10$ km s$^{-1}$ causes a significant suppression of AGN feedback. The magnitude of this suppression depends strongly on the imposed drift speed: a lower value leads to a later onset of effective AGN feedback, and hence a higher central stellar density of haloes with $M_\mathrm{200c} > 10^{12}\,\msun$, while higher values approach the outcome of instantaneous repositioning (Fig.~\ref{fig:eagle_vrepos}).
	
	\item Tying SMBHs less strongly to the centre of their host galaxies, while not disabling repositioning completely, restricts significant SMBH growth to more massive haloes. Where SMBHs can grow, they reach the state of self-regulated growth at lower masses, consistent with feedback from off-centre AGN being locally more efficient, but with a reduced impact on their host galaxy (Fig.~\ref{fig:bahamas_vdrift}).
	
	\item The effect of repositioning is not limited to moving SMBHs towards the centre of their galaxy right after they are seeded, nor to allowing them to overcome slow gas accretion at low mass. Efficient AGN feedback requires repositioning of SMBHs with $m_\mathrm{BH} \gg m_\mathrm{seed}$, especially at comparatively high resolution (Fig.~\ref{fig:growth_repos}).
	
	\item There are three mechanisms through which repositioning affects SMBH growth, and hence AGN feedback. Firstly, it enables SMBH mergers (Fig.~\ref{fig:bh_positions}) -- which lead to higher SMBH masses and hence increase the Bondi-Hoyle-Lyttleton accretion rate -- although these contribute only a minor fraction of the total mass of SMBHs with $m_\mathrm{BH} \gtrsim 10^6\,\msun$ (Fig.~\ref{fig:bh_facc}). Secondly, it moves SMBHs to regions of higher gas density, by up to several orders of magnitude. Thirdly, it (indirectly) slows SMBHs down by an order of magnitude with respect to their ambient gas. Since the Bondi-Hoyle-Lyttleton accretion rate depends on this velocity to the third power, this suppresses gas accretion, typically by an even larger factor than due to the lower ambient gas densities (Fig.~\ref{fig:bh_densvel}).
	
\end{enumerate} 

Although the impact of, and more realistic alternatives to, repositioning of SMBHs have been explored in a number of recent works based on higher-resolution simulations (e.g.~\citealt{Tremmel_et_al_2015,Pfister_et_al_2019,Ma_et_al_2021}), the discussion has so far focused mostly on the SMBHs themselves. In addition, all of these proposed schemes ignore the unresolved mass bound to the black hole. Our results show that this modelling detail has far wider-reaching consequences, comparable to the implementation of SMBH accretion or AGN feedback. Although it may be possible to account for the uncertainty associated with repositioning empirically through calibration of other SMBH model parameters, strong predictions for the growth of massive haloes and cosmic large-scale structure require an effort to enable the realistic modelling of SMBH dynamics, across a wide range of resolutions. Alternatively, progress might be made by calibrating repositioning-related parameters (e.g.~the drift speed, but also the SMBH seed mass which determines how much growth can be achieved through SMBH mergers) against observations, as is already commonly done for supernova feedback. 

\section*{Acknowledgements}

We thank the referee for helpful suggestions that have improved the presentation of our results, and Robert Grand, Klaus Dolag, Volker Springel, and Rainer Weinberger for helpful clarifications about SMBH repositioning in the \textsc{gadget} and \textsc{arepo} simulation codes.

The authors gratefully acknowledge funding from the Netherlands Organization for Scientific Research (NWO) through Veni grant number 639.041.749, Veni grant number 639.041.751, and Vici grant number 639.043.409. This work was supported by STFC consolidated grant number ST/T000244/1. JB is supported by STFC studentship ST/R504725/1. EC is supported by the funding from the European Union's Horizon 2020 research and innovation programme under the Marie Sk{\l}odowska-Curie grant agreement No 860744. This work used the DiRAC@Durham facility managed by the Institute for Computational Cosmology on behalf of the STFC DiRAC HPC Facility (www.dirac.ac.uk). The equipment was funded by BEIS capital funding via STFC capital grants ST/K00042X/1, ST/P002293/1, ST/R002371/1 and ST/S002502/1, Durham University and STFC operations grant ST/R000832/1. DiRAC is part of the National e-Infrastructure.

The research in this paper made use of the SWIFT open-source simulation code
(\url{http://www.swiftsim.com}, \citealt{Schaller_et_al_2018}) version 0.9.0. This work made use of the \texttt{python} libraries \texttt{numpy} \citep{Harris_et_al_2020}, \texttt{matplotlib} \citep{Hunter_2007}, and \texttt{swiftsimio} \citep{Borrow_Borrisov_2020}, and of NASA's Astrophysics Data System Bibliographic Services.

\section*{Data Availability}

The data shown in the plots of this paper is available at \url{https://home.strw.leidenuniv.nl/~bahe/papers.html}. The \swift{} simulation code, including the full subgrid physics model used here and the initial conditions for the \eagle{} resolution simulations, is publicly available at \url{http://www.swiftsim.com}. Simulation outputs can be obtained from the corresponding author on reasonable request.



\bibliographystyle{mnras}
\bibliography{bibliography_repositioning} 




\appendix

\section{Galaxies without any SMBH}
\label{app:bh_free_galaxies}
\begin{figure}
	\includegraphics[width=\columnwidth]{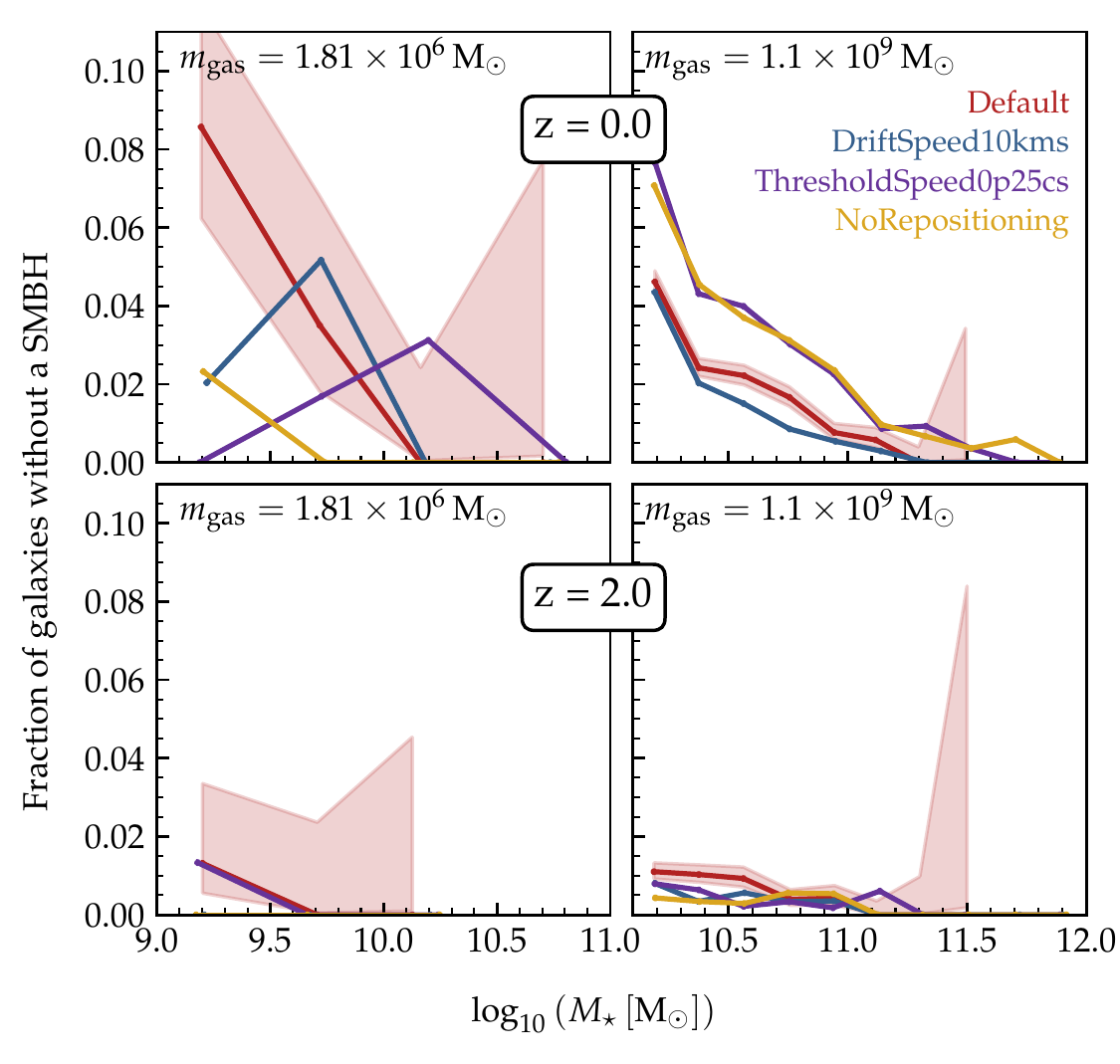}
	\caption{Fraction of galaxies that do not contain a single SMBH in four different repositioning models (different colours, see top-right panel). The top and bottom rows show results at $z = 0$ and $z = 2$, respectively, while the left and right columns are for simulations at \textsc{Eagle} and \textsc{Bahamas} resolution, respectively. Shaded bands indicate binomial $1\sigma$ uncertainties following \citet{Cameron_2011}; they are shown only for the \texttt{Default} model for clarity. At $z = 0$, the \texttt{Default} \textsc{Eagle} resolution simulation contains a modest fraction of galaxies without SMBHs at $\mstar \lesssim 10^{10}\,\msun$ due to erroneous repositioning. At higher redshift, lower resolution, or with reduced repositioning, this problem is less evident.}
	\label{fig:app_nobh}
\end{figure}

Although repositioning of SMBHs is intended to keep them at the centre of their host galaxy, during close encounters it may also lead to artificial SMBH transfers to a different galaxy. We quantify the occurrence of this in Fig.~\ref{fig:app_nobh}, where we plot the fraction of galaxies that do not host a single SMBH as a function of $\mstar$, for four different repositioning models (\texttt{Default, DriftSpeed10kms, ThresholdSpeed0p25cs, NoRepositioning}), two resolutions, and at two redshifts. We note first that SMBH-free galaxies occur even when repositioning is completely disabled (\texttt{NoRepositioning}, yellow), because some galaxies at the low-mass end of the ranges shown have not yet reached the required halo mass for SMBH seeding (especially in the lower-resolution simulations, where this threshold is higher)\footnote{We remind the reader that our SMBH seeding procedure is an extremely oversimplified model that cannot make strong predictions on the SMBH occupation fraction. While alternative prescriptions have been explored in the literature (e.g.~\citealt{Tremmel_et_al_2017}), these generally require significantly higher resolution than what is currently affordable by simulations or large representative volumes.}.
However, at \textsc{Eagle} resolution, SMBH-free galaxies are more common with the \texttt{Default} repositioning model at $z = 0$, at least at the low-mass end (close to 10 per cent at $\mstar \lesssim 10^{10}\,\msun$). Reducing the repositioning speed (\texttt{DriftSpeed10kms}, blue) or restricting repositioning to slow-moving neighbours (\texttt{ThresholdSpeed0p25cs}, purple) generally reduces the occurrence of SMBH-free galaxies, but does not prevent them completely.

At $z = 2$, their occurrence is reduced to $\approx$1 per cent in all models, indicating that SMBH loss builds up over cosmic time. In the lower-resolution simulations there is no clear excess of SMBH free galaxies in the resolved mass range; in fact the \texttt{Default} model here produces fewer such cases at fixed $\mstar$. This may seem counter-intuitive at first, but can be understood in terms of the suppressed star formation efficiency such that a smaller fraction of galaxies at a given $\mstar$ fall below the SMBH seeding halo mass threshold.
	
We point out that the repositioning-related excess of SMBH-free galaxies is driven entirely by satellites, rather than centrals. On the one hand, this follows directly from the SMBH seeding implementation in our model: if a central somehow lost its SMBH, it would quickly be replaced by a newly seeded one. However, we have verified that none of the SMBHs that are the most massive ones in their $\mstar > 10^9\,\msun$ galaxy were seeded after $z = 2$ in the \texttt{Default} \textsc{Eagle}-resolution run, which rules out the possibility of such ``covered up'' SMBH losses in centrals. Instead, we have found that SMBHs are primarily lost from satellite galaxies when they have a close pericentric passage with their central, when the SMBH is repositioned towards the central galaxy (and, typically, merged with its own SMBH), while the satellite galaxy still survives for some time.

\section{Impact of SMBH model changes from EAGLE}
\label{app:bh_changes}

As discussed in Section \ref{sec:model}, the baseline SMBH model in our simulations includes a number of updates compared to the \eagle{} model described in \citet{Schaye_et_al_2015}. To assess the impact of these changes, we have run a number of additional simulations that revert one of these changes at a time: (i) the implementation of gas transfer from neighbouring particles (run \texttt{GasSwallowing}); (ii) calculating the sound speed for gas near the entropy floor (\texttt{RawSoundSpeed}); (iii) omitting the accretion-rate based SMBH time step limiter (\texttt{NoBHTimeStepLimiter}); (iv) the SMBH merger criterion (\texttt{OldMergers}); (v) the choice of which neighbour particle(s) to heat in AGN feedback (\texttt{RandomAGN}); and (vi) the angular-momentum-based reduction of SMBH accretion rates (\texttt{WithAngMomLimiter}). 

\begin{figure}
	\includegraphics[width=\columnwidth]{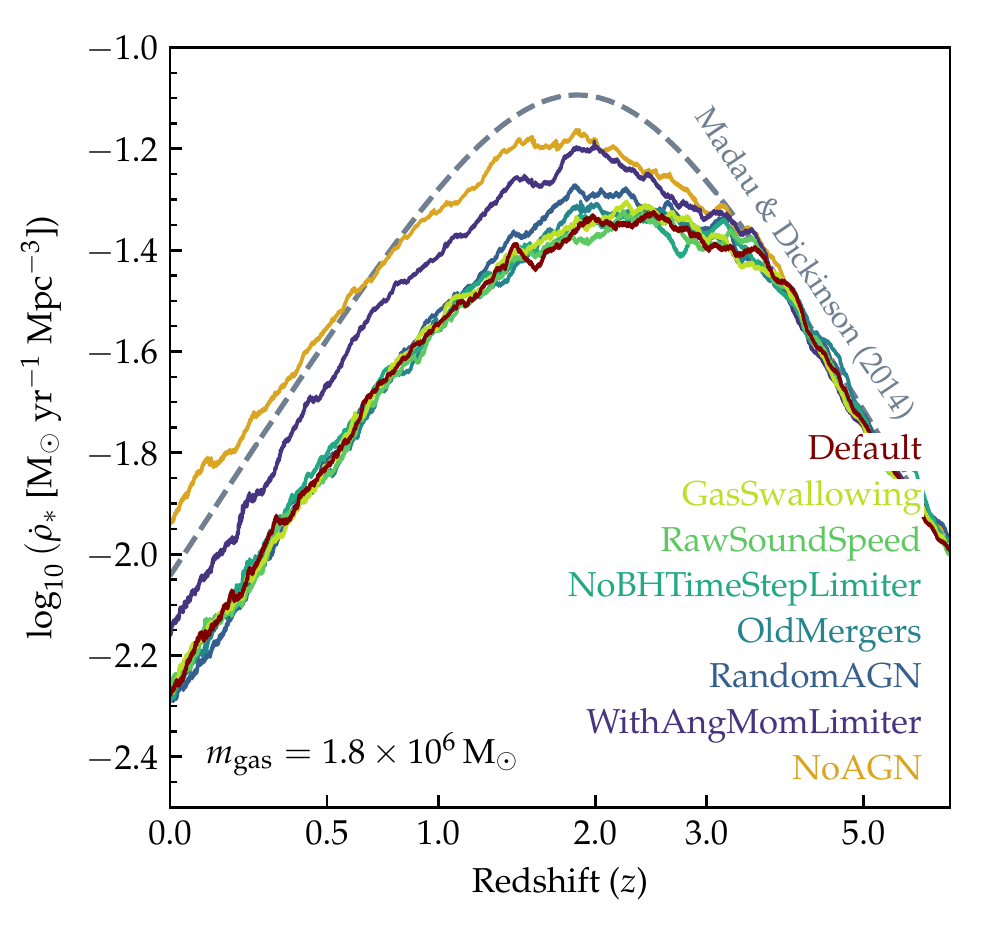}
	\caption{Global star formation rate density in \eagle{}-resolution simulations that vary individual elements of our baseline SMBH model (shades of blue/green). The baseline model itself (\texttt{Default}) and the simulation with AGN feedback disabled (\texttt{NoAGN}) are shown in red and yellow, respectively. For guidance, the dashed grey line represents the observational best fit of \citet{Madau_Dickinson_2014}; the simulations have not been calibrated to reproduce this. Besides \texttt{NoAGN}, the only variation that deviates significantly from the \texttt{Default} setup is \texttt{WithAngMomLimiter} (purple) that includes the angular momentum dependent SMBH accretion suppression term of \citet{Rosas-Guevara_et_al_2015}.}
	\label{fig:app_sfrh}
\end{figure}

In Fig.~\ref{fig:app_sfrh} we compare the evolution of the cosmic star formation rate density $\dot{\rho}_\star\, (z)$ as predicted by these variations, for the \eagle{}-resolution setup. For comparison, the \texttt{Default} and \texttt{NoAGN} models are also shown. Although there are some slight deviations between the model variations, in particular around $z = 2$, most runs track each other very closely; in other words, the changes have little impact on the global simulation outcome. The same is true for the \bahamas{}-like simulations (not shown).

The one exception is \texttt{WithAngMomLimiter} (purple), the variation that adds the angular momentum limiter term of \citet{Rosas-Guevara_et_al_2015} to the Bondi-Hoyle-Lyttleton accretion formula; with it, the global SFR is enhanced by $\approx$50 per cent at $z \lesssim 2$. This is consistent with \citet{Crain_et_al_2015}, who showed a qualitatively similar effect in runs that varied the coupling between angular momentum and the suppression factor (their fig.~11; \citealt{Valentini_et_al_2020} report similar results for their independent implementation). Suitable changes to other subgrid parameters can compensate for this difference and lead to realistic star formation histories with a standard Bondi-Hoyle-Lyttleton accretion formula.

\begin{figure}
	\includegraphics[width=\columnwidth]{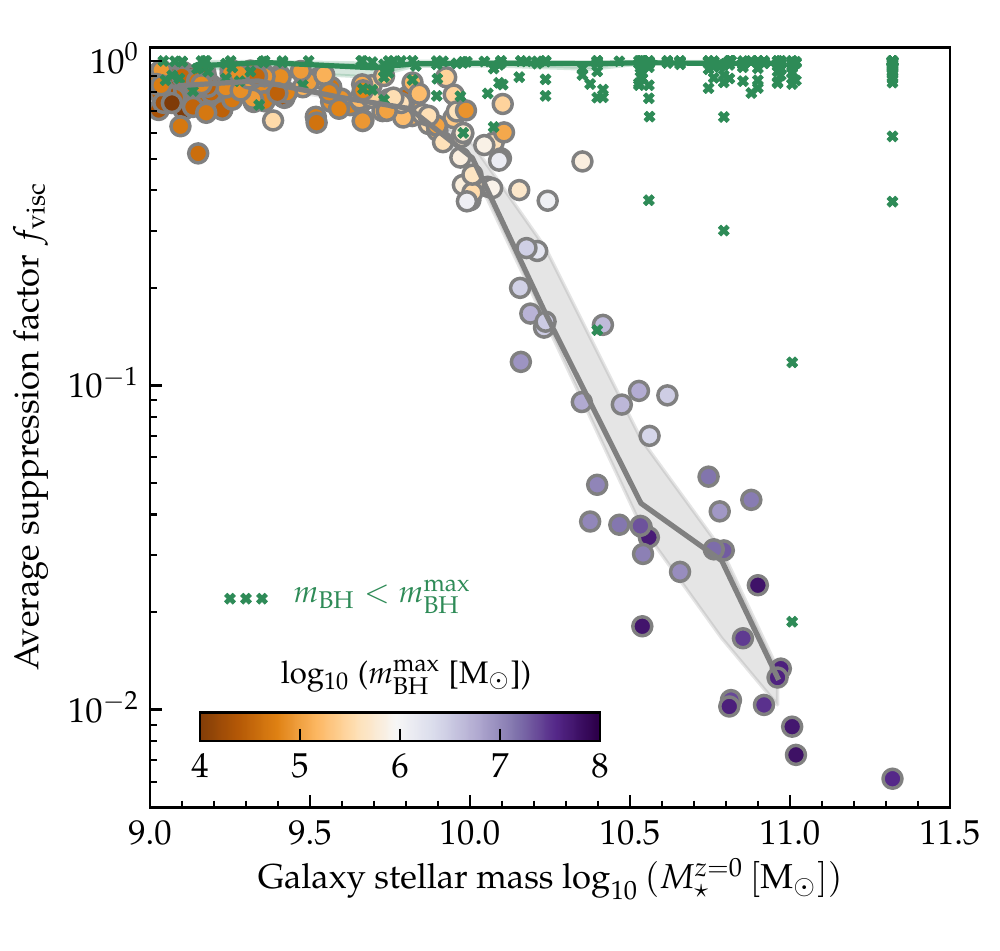}
	\caption{Dependence of the time-averaged, accretion-rate weighted SMBH accretion suppression factor $f_\mathrm{visc}$ of \citet{Rosas-Guevara_et_al_2015} on galaxy stellar mass in our model, at \eagle{} resolution. Large circles represent SMBHs that are the most massive ones in their halo, colour-coded according to their mass; the grey line and band traces their running median and $1\sigma$ uncertainty. Small green crosses represent other, subdominant SMBHs. Contrary to expectations, $f_\mathrm{visc}$ preferentially suppresses gas accretion in the dominant SMBHs of \emph{massive} galaxies in this simulation, by a factor of up to $\sim$100.}
	\label{fig:app_fvisc}
\end{figure}

In addition to its simpler nature, there is an empirical reason for preferring this approach to what was done in \eagle{}. In \citet{Rosas-Guevara_et_al_2015}, the angular momentum term was found to preferentially suppress AGN feedback in low-mass galaxies when added to the subgrid model from the OWLS project \citep{Schaye_et_al_2010}. Our simulations, however, display the opposite behaviour: as shown in Fig.~\ref{fig:app_fvisc}, the multiplicative factor $f_\mathrm{visc}$ is typically close to unity for the most massive SMBH in low-mass galaxies, but decreases systematically with stellar mass down to $\approx$\,0.01 at $\mstar = 10^{11}\,\msun$. We speculate that this difference originates from the different stellar feedback in our simulations compared to the ones of \citet{Rosas-Guevara_et_al_2015}, which has a strong impact on SMBH growth \citep{Bower_et_al_2017} but also on the structure of stars and the gas that form them \citep{Crain_et_al_2015}. It is conceivable that this leads to a higher degree of rotation in the central gas of massive galaxies, especially at $z \gg 0$, which is then perpetuated by the suppression of AGN feedback through the angular momentum based SMBH accretion limiter. Although only the result from the \eagle{} resolution run is shown in Fig.~\ref{fig:app_fvisc}, we have verified that a qualitatively similar picture emerges at \bahamas{} resolution.

\bsp	
\label{lastpage}
\end{document}